\documentclass[preprint2]{aastex631}

\usepackage{enumitem}
\accepted{\today}

\submitjournal{PASP}

\shorttitle{KMTNet Nearby Galaxy Survey}
\shortauthors{Byun et al.}
\graphicspath{{./}{figures/}}

\begin{document}

\title{KMTNet Nearby Galaxy Survey: Overview and a Survey Description}

\correspondingauthor{Minjin Kim}
\email{mkim@knu.ac.kr}

\author[0000-0002-7762-7712]{Woowon Byun}
\affiliation{Korea Astronomy and Space Science Institute, Daejeon 34055, Republic of Korea}
\affiliation{University of Science and Technology, Korea, Daejeon 34113, Republic of Korea}

\author[0000-0002-3211-9431]{Yun-Kyeong Sheen}
\affiliation{Korea Astronomy and Space Science Institute, Daejeon 34055, Republic of Korea}

\author[0000-0001-9561-8134]{Kwang-Il Seon}
\affiliation{Korea Astronomy and Space Science Institute, Daejeon 34055, Republic of Korea}
\affiliation{University of Science and Technology, Korea, Daejeon 34113, Republic of Korea}

\author[0000-0001-6947-5846]{Luis C. Ho}
\affiliation{Kavli Institute for Astronomy and Astrophysics, Peking University, Beijing 100871, People's Republic of China}
\affiliation{Department of Astronomy, School of Physics, Peking University, Beijing 100871, People's Republic of China}

\author[0000-0003-3451-0925]{Joon Hyeop Lee}
\affiliation{Korea Astronomy and Space Science Institute, Daejeon 34055, Republic of Korea}

\author[0000-0002-0145-9556]{Hyunjin Jeong}
\affiliation{Korea Astronomy and Space Science Institute, Daejeon 34055, Republic of Korea}

\author[0000-0001-9670-1546]{Sang Chul Kim}
\affiliation{Korea Astronomy and Space Science Institute, Daejeon 34055, Republic of Korea}
\affiliation{University of Science and Technology, Korea, Daejeon 34113, Republic of Korea}

\author[0000-0002-6982-7722]{Byeong-Gon Park}
\affiliation{Korea Astronomy and Space Science Institute, Daejeon 34055, Republic of Korea}

\author[0000-0001-7594-8072]{Yongseok Lee}
\affiliation{Korea Astronomy and Space Science Institute, Daejeon 34055, Republic of Korea}
\affiliation{School of Space Research, Kyung Hee University, Yongin, Kyeonggi 17104, Republic of Korea}

\author[0000-0002-7511-2950]{Sang-Mok Cha}
\affiliation{Korea Astronomy and Space Science Institute, Daejeon 34055, Republic of Korea}
\affiliation{School of Space Research, Kyung Hee University, Yongin, Kyeonggi 17104, Republic of Korea}


\author[0000-0002-3560-0781]{Minjin Kim}
\affiliation{Department of Astronomy and Atmospheric Sciences, Kyungpook National University, Daegu 41566, Republic of Korea}



\begin{abstract}

Recently, there has been an increasing demand for deep imaging surveys to investigate the history of the mass assembly of galaxies in detail by examining the remnants of mergers and accretions, both of which have very low surface brightness (LSB). In addition, the nature of star formation in LSB regions, such as galaxy outer disks, is also an intriguing topic in terms of understanding the physical mechanisms of disk evolution. To address these issues, this study conducts a survey project, called the Korea Microlensing Telescope Network (KMTNet) Nearby Galaxy Survey to construct a deep imaging data set of nearby galaxies in the southern hemisphere using KMTNet. It provides deep and wide-field images with a field-of-view of $\sim$12 deg$^2$ for 13 nearby galaxies drawn from the Carnegie-Irvine Galaxy Survey catalog, in optical broadbands ($BRI$) and an H$\alpha$ narrowband. Through a dedicated data reduction, the surface brightness limit in 10$^{\prime\prime}\times10^{\prime\prime}$ boxes was found to reach as deep as $\mu_{1\sigma}\sim29$--31 mag arcsec$^{-2}$ in the optical broadbands and $f_{1\sigma}\sim1$--$2\times 10^{-18}$ erg s$^{-1}$ cm$^{-2}$ arcsec$^{-2}$ in the H$\alpha$ narrowband. To conclude the paper, several possible scientific applications for this data set are described.

\end{abstract}



\section{Introduction} \label{sec:intro}
In the $\Lambda$CDM model, galaxies are formed and evolve in a hierarchical manner, in the sense that they grow through mergers and the smooth accretion of satellite galaxies \citep[e.g.,][]{1991ApJ...379...52W}. In this model, low surface brightness (LSB) features (e.g., stellar halos, extended star formations, and dwarf galaxies) in the galaxy outskirts can be used to probe in detail the history of mass assembly in the host galaxies \citep[e.g.,][]{2005ApJ...635..931B}. Therefore, studying the physical properties of these LSB features will enable a better understanding of the recent evolution of the corresponding galaxies. However, it is technically challenging to study LSB features observationally due to their relative faintness compared with the underlying sky brightness. For example, reaching a surface brightness of 27--28 mag arcsec$^{-2}$ requires suppression of the sky fluctuation down to 0.5\%. In addition, contamination from the point-spread function as well as scattered light and dust in Galactic cirrus are non-negligible at this level of depth \citep[e.g.,][]{2011ApJ...731...89T,2015MNRAS.446..120D,2022ApJ...932...44G}. 

Recently, several survey projects have been carried out to study LSB features in nearby galaxies. For example, the Dragonfly Nearby Galaxies Survey obtained deep and wide-field images of nearby galaxies using the Dragonfly Telephoto Array, reaching up to 30--31 mag arcsec$^{-2}$ \citep{2014PASP..126...55A}. Based on this survey, \cite{2016ApJ...833..168M} and \cite{2018ApJ...855...78Z} demonstrated that the stellar halo fractions among the spiral galaxies are diverse, indicating that their formation history is non-uniform \citep[see also][]{2022ApJ...932...44G}. In addition, a large number of dwarf galaxies and ultra-diffuse galaxies have been identified by this survey, and further studies have been extensively carried out to unveil their physical nature \citep[e.g.,][]{2015ApJ...798L..45V,2016ApJ...828L...6V,2017ApJ...844L..11V,2018ApJ...856...69D,2018ApJ...859...37G,2022Natur.605..435V}. The great success of this unprecedented deep survey clearly reveals that deep images of nearby galaxies are of importance in understanding their evolutionary history. 

Dwarf galaxies in particular can serve as key probes to test the $\Lambda$CDM model on a small scale. The observed number of dwarf satellite galaxies is substantially lower than that predicted by the theoretical studies, an issue referred to as the missing satellites problem \citep[e.g.,][]{1999ApJ...522...82K,1999ApJ...524L..19M}. To reconcile this discrepancy, several scenarios have been proposed. One solution suggests that dark matter halos are composed of warm dark matter rather than cold dark matter, helping to partly prevent the formation of small scale structures \citep[e.g.,][]{2001ApJ...556...93B,2019ApJ...878L..32N}. One alternative solution is that star formation (SF) in dwarf galaxies is simply suppressed due to the ultraviolet (UV) background or supernova explosions \citep[e.g.,][]{1997ApJ...490..493N,2003MNRAS.344.1131D,2020MNRAS.491.1656A}. Finally, it has long been suggested that this issue may arise from observational incompleteness \citep[e.g.,][]{2007ApJ...656L..13I,2007ApJ...670..313S,2020MNRAS.493.2596F}, as a substantial number of ultra-faint dwarf galaxies have recently been discovered. However, previous studies have mostly relied on the observational results from the Local Group (i.e., our Galaxy and M31). Therefore, from an observational point of view, it is desirable to search for the faint satellite galaxies of various types beyond the Local Group using deep imaging. 

One of the intriguing characteristics in the outskirts of nearby galaxies is that the SF properties are clearly distinct from those in the inner regions. For example, based on the Galaxy Evolution Explorer (GALEX) mission, \cite{2005ApJ...619L..79T,2007ApJS..173..538T} reported a significant fraction of nearby disk galaxies exhibiting a diffuse extended UV emission in their outer disks. These are known as extended UV (XUV)-disk galaxies. This indicates that SF actively occurs in the outer disk, and may serve as indirect evidence for inside-out disk formation \citep[e.g.,][]{2021ApJ...923..199P}. While the physical origin of this feature is still under debate \citep[e.g.,][]{2005ApJ...627L..29G,2008AJ....136..479D,2019ApJ...881...71E,2021JApA...42...85D}, the flux ratio of H$\alpha$ to UV in the XUV disks appears to be lower than that in typical star-forming regions \citep[e.g.,][]{2007AJ....134..135Z,2010MNRAS.405.2791G}.
A similar discrepancy between UV and H$\alpha$ has also been reported in dwarf galaxies \citep[e.g.,][]{2009ApJ...706..599L}. Such an H$\alpha$ deficit in low-stellar-density regions can be attributed to poor sampling of the initial mass function \citep[IMF; e.g.,][]{2011ApJ...741L..26F,2014MNRAS.444.3275D}, a non-uniform IMF \citep[e.g.,][]{2009ApJ...695..765M,2009MNRAS.395..394P,2018A&A...620A..39J, 2020MNRAS.491.2366B, 2021MNRAS.506.4979K}, a leakage of UV photons \citep[e.g.,][]{1997MNRAS.291..827O,2011MNRAS.411..235E,2012MNRAS.423.2933R,2020ApJ...902...54C}, and a diverse SF history \citep[SFH; e.g.,][]{2004A&A...421..887I,2004MNRAS.350...21S,2012ApJ...744...44W,2019ApJ...881...71E}. A detailed analysis of the stellar population and SFH in XUV disks, estimated from deep multi-band images and deep H$\alpha$ images, is crucial to exploring the SF mechanism in XUV disks. 

To extensively study the LSB features in nearby galaxies, this study conducts a deep and wide-field imaging survey of nearby galaxies from the southern hemisphere using the Korea Microlensing Telescope Network \cite[KMTNet;][]{2016JKAS...49...37K}. In this paper, an overview of the survey (the KMTNet Nearby Galaxy Survey; KNGS) is presented. In Section \ref{sec:obs}, the KMTNet is briefly introduced and the sample selection and observation strategy are described. A detailed data reduction process used to minimize the sky fluctuation is presented in Section \ref{sec:datared}, and an assessment of the imaging quality is described in Section \ref{sec:iq}. Finally, the potential scientific applications of the survey are summarized in Section \ref{sec:app}. 

\begin{deluxetable*}{lcccccccccc}[t]
\tabletypesize{\footnotesize}
\tablenum{1}
\tablecaption{Information on the Samples of KMTNet Nearby Galaxy Survey \label{tab:sample}}
\tablewidth{0pt}
\tablehead{
\colhead{Name} & \colhead{R.A.$^\mathrm{a}$} & \colhead{Decl.$^\mathrm{a}$} & \colhead{Distance$^\mathrm{a}$} & \colhead{$m_B$$^\mathrm{a}$} & \colhead{Morphology$^\mathrm{a}$} & $t_B^\mathrm{tot}$ & $t_R^\mathrm{tot}$ & $t_I^\mathrm{tot}$ & $t_\mathrm{H\alpha}^\mathrm{tot}$ & Note \\
& \colhead{(hh:mm:ss)} & \colhead{(dd:mm:ss)} & \colhead{(Mpc)} & \colhead{(mag)} & & (hr) & (hr) & (hr) & (hr) &
}
\startdata
NGC 1097 & 02:46:19.05 & $-$30:16:29.6 & 16.8 & 10.0 & SB(s)b & 4.4 & 2.1 & 3.6 & 3.2 & XUV$^\mathrm{b}$ \\
NGC 1291 & 03:17:18.59 & $-$41:06:29.0 & 9.1 & 9.4 & (R)SB(s)0/a & 2.8 & 1.6 & 1.6 & 2.1 & Peculiar$^\mathrm{b}$  \\
NGC 1316 & 03:22:41.72 & $-$37:12:29.6 & 17.5 & 9.4 & SAB(s)0 & 4.7 & 4.9 & 4.3 & 2.5 &  \\
NGC 1512 & 04:03:54.28 & $-$43:20:55.9 & 10.4 & 11.3 & SB(r)a & 4.8 & 5.0 & 4.6 & 1.9 & XUV \\
NGC 1672 & 04:45:42.50 & $-$59:14:49.9 & 11.4 & 10.8 & SB(s)b & 6.6 & 4.4 & 5.2 & 1.8 & XUV \\
NGC 2090 & 05:47:01.89 & $-$34:15:02.2 & 11.3 & 11.8 & SA(rs)c & 4.6 & 4.9 & 4.7 & 1.9 & XUV \\
ESO 556-012 & 06:17:49.24 & $-$21:03:38.0 & 49.9 & 14.9 & SB(s)m & 4.8 & 4.7 & 4.1 & - & XUV \\
ESO 208-021 & 07:33:56.25 & $-$50:26:34.9 & 13.3 & 12.0 & SAB0 & 4.5 & 4.7 & 4.7 & 1.8 & \\
NGC 2784 & 09:12:19.50 & $-$24:10:21.4 & 9.6 & 11.2 & SA0(s) & 4.3 & 3.1$^\mathrm{c}$ & 3.8 & 2.6 & \\
NGC 2835 & 09:17:52.91 & $-$22:21:16.8 & 8.8 & 11.1 & SB(rs)c & 3.7$^\mathrm{c}$ & 3.1 & 3.8 & 0.7 & \\
NGC 3621 & 11:18:16.51 & $-$32:48:50.6 & 6.6 & 9.7 & SA(s)d & 7.0 & 4.6 & 4.5 & 3.5 & XUV \\
NGC 3923 & 11:51:01.69 & $-$28:48:21.7 & 21.3 & 10.9 & E4-5 & 2.9 & 5.1 & 4.1 & 1.3 & \\
NGC 5236 & 13:37:00.95 & $-$29:51:55.5 & 4.6 & 8.3 & SAB(s)c & 5.4 & 4.5 & 4.7 & 3.7 & XUV \\
\enddata
\tablecomments{The total exposure time was calculated from only the images used to create the final mosaic image.\\
$^\mathrm{a}$ NASA Extragalactic Database (NED) \\
$^\mathrm{b}$ This was additionally classified by \cite{2010MNRAS.405.2791G}. \\
$^\mathrm{c}$ The chip gap is not completely filled due to the observing condition.}
\end{deluxetable*}

\section{Sample and Observation} \label{sec:obs}
\subsection{Sample}
The KNGS aims to take deep optical images of nearby galaxies to explore faint structures, which may provide crucial hints on the history of these galaxies' evolution. This study's primary sample of nearby galaxies was adopted from the Carnegie-Irvine Galaxy Survey \citep[CGS;][]{2011ApJS..197...21H,2011ApJS..197...22L}. The CGS performed optical broadband imaging of a sample of 605 bright southern galaxies using the du Pont 2.5-m telescope at Las Campanas Observatory. The KMTNet was able to revisit the target galaxies using new images with a field-of-view (FoV) of $2^\circ\times2^\circ$, wider than the small FoV of the CGS ($8.^{\prime}9\times8.^{\prime}9$). Even when observing nearby galaxies, the large FoV enables greater dithering and secures many more pixels in the sky area, thereby guaranteeing a more robust estimation of the background sky. This allowed us to perform flat-fielding using a dark-sky flat and individual sky subtraction (see Section \ref{sec:datared}). Indeed, photometric uncertainty can be introduced by many factors, such as the detector, optics, or even data processing. This study's strategy took advantage of a wide FoV to minimize the uncertainty arising from improper data reduction. 

While the CGS catalog includes galaxies with a wide range of morphological types, this study focused on the outskirts of spiral galaxies with LSB features. One of the intriguing types of targets was galaxies with an XUV-disk beyond their optical extent. These XUV disks indicate recent SF at a low surface density at the outskirts of the galaxy. \cite{2007ApJS..173..538T} searched XUV-disk galaxies using the far-UV and near-UV images of the GALEX satellite \citep{2005ApJ...619L...1M}. This study matched the CGS catalog with the XUV-disk galaxy list and prioritized these galaxies in the survey. To maximize the merit of the KMTNet's wide FoV, we preferentially selected observation targets located at a close distance. Most of the target galaxies were within 20 Mpc, with ESO 556-012 being the farthest away at 49.9 Mpc. Additionally, a control sample of non-XUV disks was selected from the CGS catalog with comparable morphological types to the XUV-disk targets. Based on a comparison between the XUV disks and non-XUV disks, the aim was to unveil the physical nature of the XUV disks. The basic information regarding the target galaxies is presented in Table \ref{tab:sample}. We make the final mosaic images publicly available on the website: \url{https://data.kasi.re.kr/vo/KNGS/}. 

\begin{figure}[t]
\centering
\includegraphics[width=80mm]{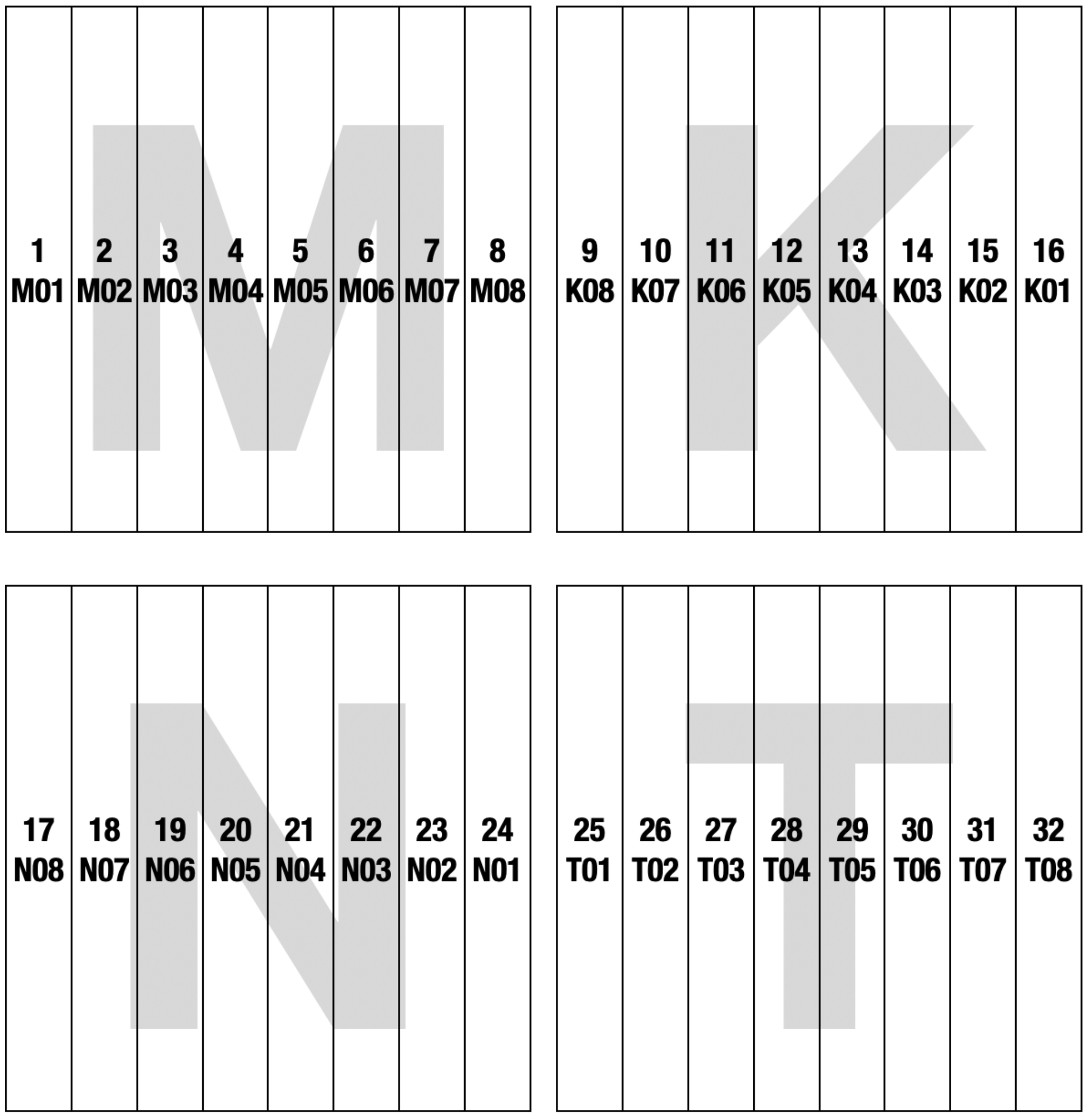}
\caption{Configuration of the KMTNet CCD imager. It consists of four $9k\times9k$ chips labeled M, K, N, and T, each of which is divided into eight amplifiers. The order and name of each amplifier are noted. \label{fig:fig1}}
\end{figure}

\subsection{Observations}
The KNGS was conducted from 2015 to 2020 using the KMTNet, consisting of three 1.6-m telescopes located at the Cerro Tololo Inter-American Observatory (CTIO), South African Astronomical Observatory, and Siding Spring Observatory. Each telescope is equipped with a mosaic CCD camera with a wide FoV of $2^\circ\times2^\circ$ and a pixel scale of 0.$^{\prime\prime}$4 pixel$^{-1}$. Figure \ref{fig:fig1} illustrates the configuration of the KMTNet CCD imager, which includes four CCD chips with eight amplifiers each. \cite{2018AJ....156..249B} have described the telescopes in more detail.

Optical images of the target galaxies were taken using $B$, $R$, $I$, and H$\alpha$ filters with the same exposure time of 120 sec for each image. For each observation, we placed the target galaxy in the center of a chip and conducted 7-point dithering to fill the gap between the four chips.\footnote{In the case of NGC 2784 and NGC 2835, the chip gap is not filled properly due to the observing condition.} Additionally, we adopted a 4-point dithering sequence with a large shift ($\sim 1$ deg) to place the target in the order of M-K-N-T in a different chip. This allowed us to use the images of the blank sky to robustly model the global background. Primarily, observations were coordinated to obtain total exposure times of $\sim$4 hr in the $BRI$ bands and $\sim$2 hr in the H$\alpha$ band for each target by the end of the survey. The H$\alpha$ narrowband filter had a bandwidth of $\sim$80 $\mathrm{\AA}$ centered at $\sim$6,570 $\mathrm{\AA}$, allowing for the detection of the H$\alpha$ emission from our targets within 30 Mpc. The total integration times of the final deep images of the $B$, $R$, $I$, and H$\alpha$ bands are listed in Table \ref{tab:sample}. According to the observation strategy, data were taken using the same telescope for the same filter. This way, we could avoid any issues resulting from slightly different observation conditions among the three KMTNet telescopes and achieve more robust calibrations of the images for co-addition. As the H$\alpha$ narrowband filter was only available in the telescope at the CTIO, all the H$\alpha$ data were obtained at the KMTNet-CTIO site. 

\section{Data Reduction} \label{sec:datared}

To fully exploit the potential of the wide-field images, we had to pay attention to the complex steps of the data reduction procedure, especially as they included flat-fielding, sky subtraction, and the determination of background uncertainties. One of the most important things we aimed to achieve was to simplify the data reduction procedure as much as possible to prevent the introduction of any additional photometric uncertainty during the data processing. As shown in Figure \ref{fig:fig1}, the configuration of the KMTNet imager is not optimal for the observation of the LSB features. Although the composition of the amplifiers enabled us to read out large amounts of imaging data faster, it introduced discontinuities into the background level among the amplifiers. This led us to search for a method to effectively remove the amplifier pattern \citep[see details in][]{2018AJ....156..249B}. Finally, we built an optimal data reduction procedure, including the steps outlined below:\footnote{The numerical results in this section are based on the $R$ band images of NGC 1291. The values can vary depending on the targets and bands, but the overall trends do not change.}

\begin{itemize}
\item{\textit{Image sampling} -- 
It is desirable that all the object frames that will be co-added have similar photometric conditions. Therefore, a strict initial sampling of the raw images was required. For instance, we discarded images with high background levels due to clouds, artifacts caused by telescope tracking errors, or abnormal bias levels resulting from some of the detectors. In addition, we used the data only if at least 15 object frames per day were guaranteed so that a dark-sky flat of the day could be created.}

\begin{figure}[t]
\centering
\includegraphics[width=80mm]{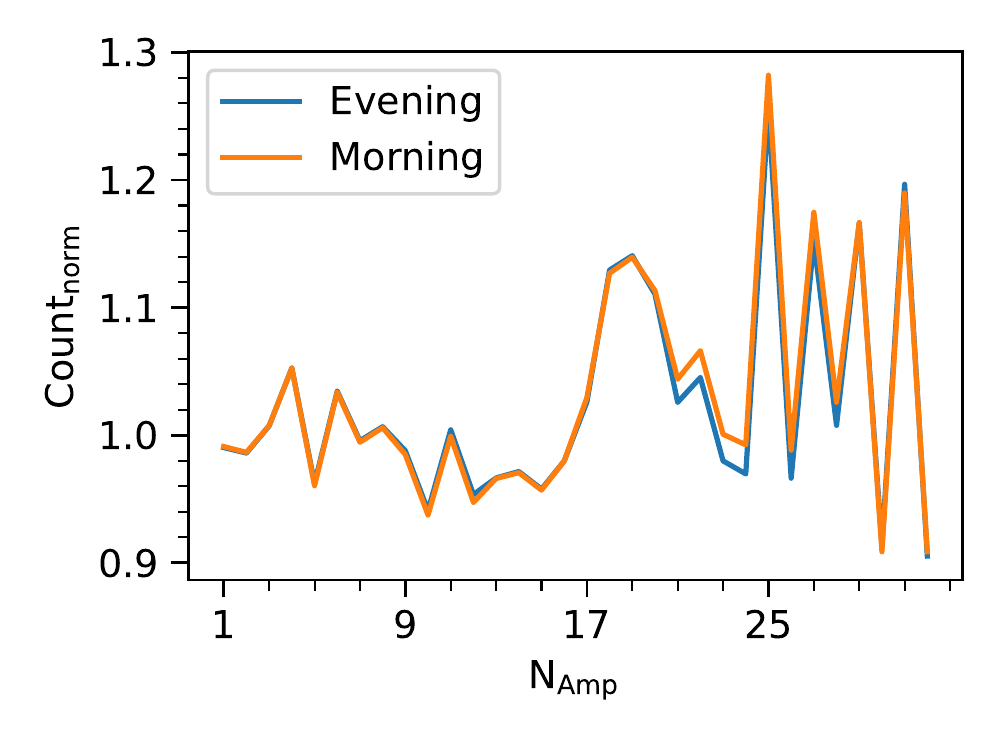}
\caption{Variations of bias level for the 32 amplifiers. The two lines represent the average trend of the evening (blue) and morning (orange), respectively, and are normalized to median values for each. The variation trends are not constant from amplifier to amplifier. \label{fig:fig2}}
\end{figure}

\item{\textit{Overscan correction and trimming} --
The bias levels appeared to change up to $\pm1$\% over a single night (Figure \ref{fig:fig2}), indicating that applying bias correction using a single master bias could be unreliable. Because we verified that the overscan followed the bias variation well (see Figure 2 in \citealt{2018AJ....156..249B}), we used overscan instead of bias for each amplifier. Note that we did not perform dark subtraction because the dark levels were revealed to be negligible.}

\item{\textit{Flat-field correction} -- 
We found that a master flat generated from twilight flat frames was inappropriate for flat-field correction. Between the evening and the morning, the level of the twilight flat changed significantly by $\pm5$\%, and the trend even reversed in some amplifiers, possibly due to the intrinsic sky gradient (Figure \ref{fig:fig3}). Consequently, a dark-sky flat was created using overscan-subtracted object frames with intensive object masking.\footnote{In fact, the sky level of the object frames also changed over time but it was less exaggerated than that of twilight frames, thereby providing a more appropriate flat image.} Object masking was carried out in two stages: a conservative detection with a sigma threshold of $\sim$1.5--3$\sigma$ using \texttt{objmasks} in IRAF and an additional detection with \texttt{Photutils}\footnote{\url{https://photutils.readthedocs.io/en/stable/}} from the Python library to mask out the diffuse light of bright objects. Although a large fraction of the pixels in each image were masked, valid pixels representing the blank sky were secured by the significant dithering and chip-by-chip movement included in our observational strategy. Flat-field correction with a dark-sky flat effectively removed discontinuity among the background levels between the amplifiers in individual images, leaving only an intrinsic sky gradient with a peak-to-peak deviation of $\sim$3\% compared with the original sky level. Meanwhile, bad pixel correction was applied to each image using \texttt{fixpix} in IRAF with the bad pixel mask generated from the dark-sky flat.}

\begin{figure}[t]
\centering
\includegraphics[width=80mm]{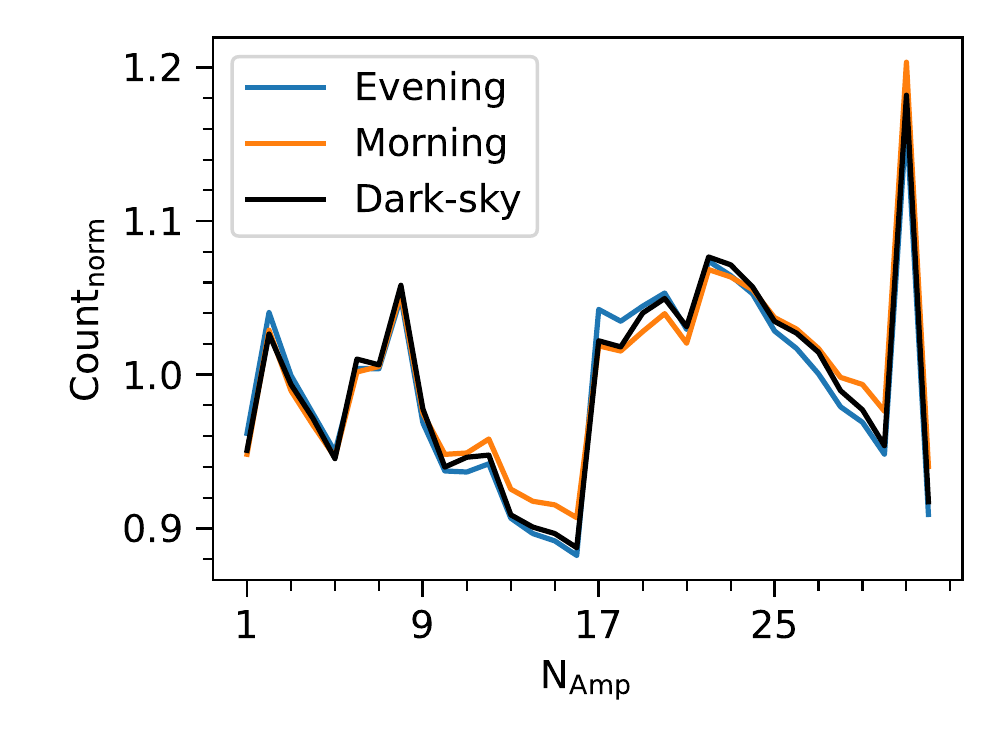}
\caption{Variations in flat level for the 32 amplifiers. The other descriptions are the same as in Figure \ref{fig:fig2}, except for the addition of a dark-sky flat (black). This shows that the trend of the dark-sky flat, which yielded the best results for the flat-field correction, is not at all consistent with either the trend of the evening or morning flats. \label{fig:fig3}}
\end{figure}

\begin{figure*}[t]
\centering
\includegraphics[width=180mm]{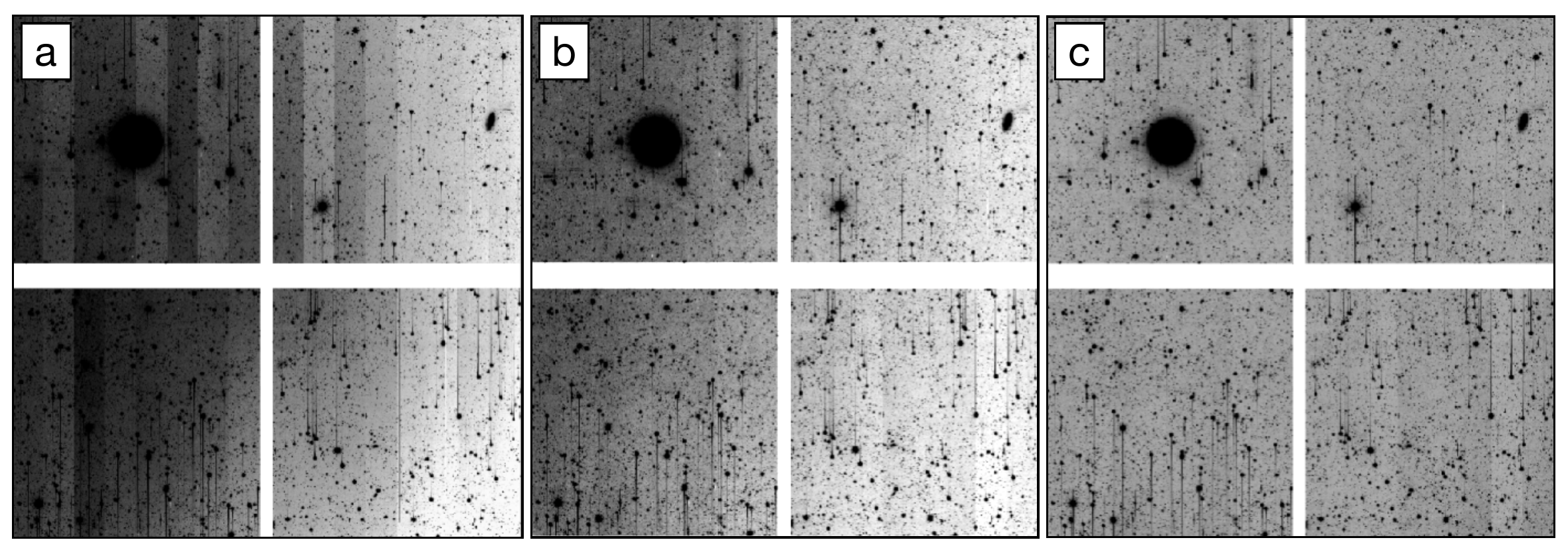}
\caption{Comparisons of results for different data reduction procedures: (a) overscan and twilight flat-field corrections, (b) overscan and dark-sky flat-field corrections, and (c) overscan and dark-sky flat-field corrections and sky subtraction. Flat-field correction using the dark-sky flat appears to be an appropriate process that allows for accurate modeling of the sky background by eliminating the amplifier pattern. The large object on the top-left is NGC 1291. \label{fig:fig4}}
\end{figure*}

\item{\textit{Sky subtraction} --
The co-addition of images with different sky gradients introduces significant photometric uncertainty. Therefore, careful sky subtraction by fitting the background before the co-addition was necessary. First, we merged each object frame with 32 amplifiers into a single image. In the same manner, the mask frames, obtained through improved processing after flat-fielding, were also merged. Second, we carried out median binning for the sky background of each object image with 500$\times$500 pixels to discard any unmasked hot pixels and diffused light around bright sources. Last, we modeled a sky background using a two-dimensional (2D) polynomial fit. Here, we adopted a quadratic function because a sky model fitted to higher-order functions can be sensitive to substructure rather than the global sky gradient. As a result of sky subtraction, the peak-to-peak deviation in the single image was reduced to less than 1\% of the original sky levels. }

\item{\textit{Co-addition} --
Astrometric calibration was conducted separately in each eight-amplifier chip using SExtractor \citep{1996A&AS..117..393B} and SCAMP \citep{2006ASPC..351..112B}. The details on this process can be found at \url{http://kmtnet.kasi.re.kr}. Finally, all the images were median-combined using SWarp \citep{2002ASPC..281..228B}, providing an FoV of $\sim$ 12 deg$^2$, which was wide enough to cover the virial radii of the target galaxies in our sample.}
\end{itemize}

\section{Imaging Quality} \label{sec:iq}
\subsection{Estimates of surface brightness limit}
This section describes the surface brightness limit of the processed data. While we were able to effectively remove the amplifier pattern and flatten the individual image to the level of $\sim$1\% of the sky value (Figure \ref{fig:fig4}), the local sky fluctuation remained in the co-added images. This background fluctuation could interfere with the accurate determination of sky level as well as the detection of LSB features. 

\begin{figure*}[t]
\centering
\includegraphics[width=180mm]{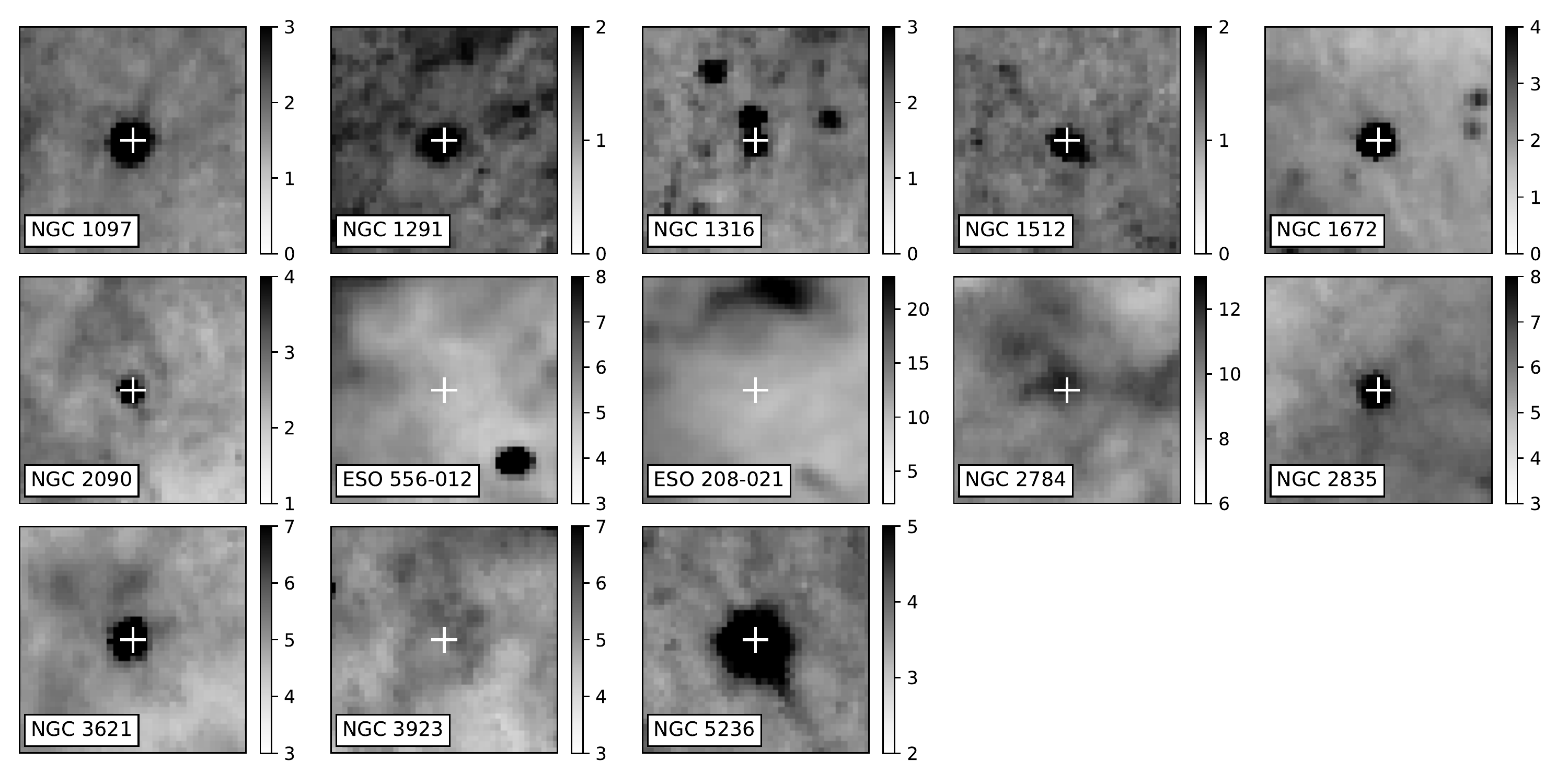}
\caption{IRAS 100$\mu$m maps with an FoV of $\sim1\times1$ deg$^2$ around the sample galaxies. The names and positions (cross) of the galaxies are denoted in each panel. The grayscale is limited by $\pm5\sigma$ of the background deviation of each image to emphasize the local fluctuation, and the units are MJy sr$^{-1}$. \label{fig:fig5}}
\end{figure*}

Since we carried out the sky subtraction by adopting a second-order polynomial to effectively fit the global sky gradient, high-order fluctuation may be caused by Galactic cirrus. In general, the Galactic dust distribution can be traced by far-infrared (FIR) measurement, in which certain images with high Galactic cirrus ($F_{100 \mu{\rm m}} \ge$ 1--2 MJy sr$^{-1}$) may be severely affected. Figure \ref{fig:fig5} shows the Infrared Astronomical Satellite (IRAS) 100$\mu$m maps around the sample galaxies with an FoV of $\sim1$ deg$^2$. These maps include fine structures and non-negligible amounts of gradient in the cirrus, which could cause non-uniform background features. Further, saturation trails and scattered light from nearby bright stars may also contribute to the fluctuation in crowded regions (see the figures in the Appendix). Therefore, the practical surface brightness limit might vary depending on the size and location of the target. 

\begin{deluxetable*}{lccccccc}[t]
\tabletypesize{\small}
\tablenum{2}
\tablecaption{The galactic coordinates and the 1$\sigma$ depths of the samples \label{tab:depth}}
\tablewidth{0pt}
\tablehead{
\colhead{Name} & \colhead{$l^\mathrm{a}$} & \colhead{$b^\mathrm{a}$} & \colhead{$\langle S_{100\mu m}\rangle^\mathrm{b}$} & \colhead{$\mu_{B,1\sigma}$} & \colhead{$\mu_{R,1\sigma}$} & \colhead{$\mu_{I,1\sigma}$} & \colhead{$f_\mathrm{H\alpha,1\sigma}$} \\
 & \colhead{\scriptsize{(deg)}} & \colhead{\scriptsize{(deg)}} & \colhead{\scriptsize{(MJy sr$^{-2}$)}} & \colhead{\scriptsize{(mag arcsec$^{-2}$)}} & \colhead{\scriptsize{(mag arcsec$^{-2}$)}} & \colhead{\scriptsize{(mag arcsec$^{-2}$)}} & \colhead{\scriptsize{(erg s$^{-1}$ cm$^{-2}$ arcsec$^{-2}$)}}
}
\startdata
NGC 1097 & 226.9 & $-$64.7 & 1.84 & 31.31 & 30.37 & 29.73 & $1.03\times10^{-18}$ \\
NGC 1291 & 247.5 & $-$57.0 & 1.51 & 31.12 & 30.13 & 29.31 & $1.29\times10^{-18}$ \\
NGC 1316 & 240.2 & $-$56.7 & 1.78 & 31.20 & 30.63 & 29.40 & $9.34\times10^{-19}$ \\
NGC 1512 & 248.7 & $-$48.2 & 1.29 & 31.38 & 30.84 & 29.85 & $1.15\times10^{-18}$ \\
NGC 1672 & 268.8 & $-$39.0 & 2.05 & 31.41 & 30.62 & 29.79 & $1.41\times10^{-18}$ \\
NGC 2090 & 239.5 & $-$27.4 & 2.59 & 30.65 & 30.75 & 29.69 & $1.23\times10^{-18}$ \\
ESO 556-012 & 228.5 & $-$16.6 & 5.41 & 31.19 & 30.49 & 29.64 &  - \\
ESO 208-021 & 262.7 & $-$14.3 & 12.40 & 30.91 & 29.87 & 29.39 & $2.30\times10^{-18}$ \\
NGC 2784 & 252.0 & 16.4 & 10.10 & 30.93 & 30.22 & 29.65 & $1.20\times10^{-18}$ \\
NGC 2835 & 251.4 & 18.5 & 6.03 & 31.17 & 30.41 & 29.72 & $6.98\times10^{-18}$ \\
NGC 3621 & 281.2 & 26.1 & 5.00 & 31.52 & 30.66 & 29.75 & $1.04\times10^{-18}$ \\
NGC 3923 & 287.3 & 32.2 & 5.12 & 30.89 & 30.73 & 29.52 & $1.19\times10^{-18}$ \\
NGC 5236 & 314.6 & 32.0 & 3.76 & 31.18 & 30.64 & 29.73 & $1.01\times10^{-18}$ \\
\enddata
\tablecomments{The calculated 1$\sigma$ depth of each band was scaled for an angular scale of $10\times10$ arcsec boxes. \\
$^\mathrm{a}$ NASA Extragalactic Database (NED) \\
$^\mathrm{b}$ Average flux of the background of $\sim$1$\times$1 deg$^2$ images from NASA/IPAC Infrared Science Archive (IRSA) \\
}
\end{deluxetable*}

\begin{figure}[t]
\centering
\includegraphics[width=80mm]{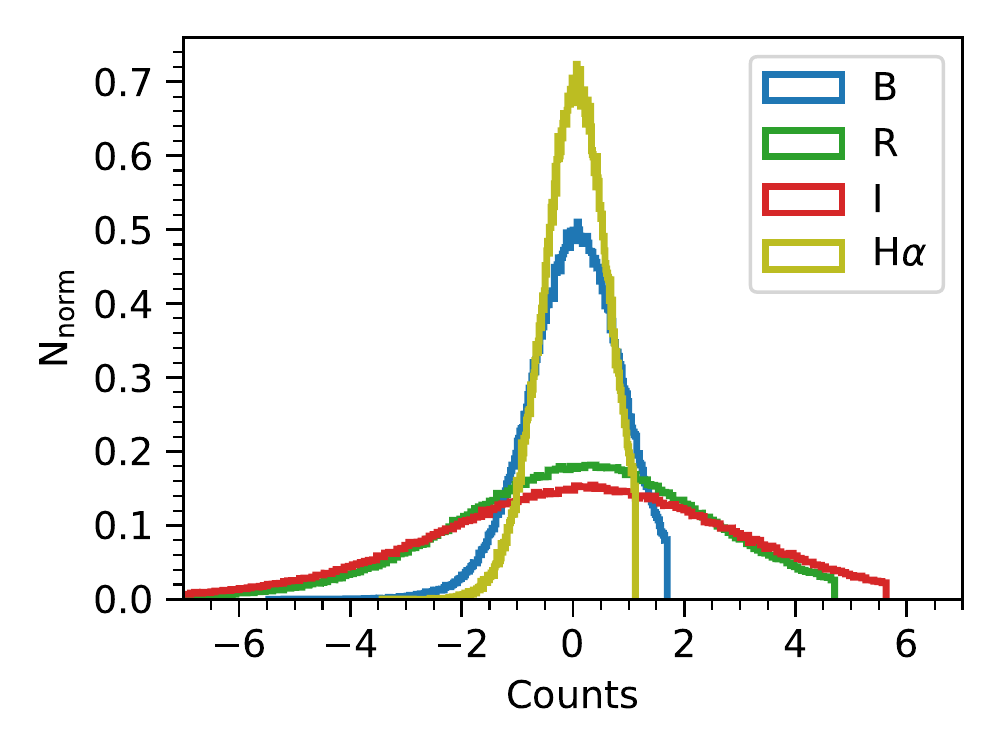}
\caption{Distribution of background values for an object-masked image around NGC 1291 with a $\sim$$2.5\times2.5$ deg$^2$ region. The histograms follow a Gaussian distribution. The upper wings are clipped out to mask bright objects. \label{fig:fig6}}
\end{figure}

To more straightforwardly compare the imaging quality of the data set, we present the surface brightness limit derived from the total background noise, which was dominated by the Poisson noise of the data. Due to the dithering pattern moving between the CCD chips, the borders of the co-added images with an FoV of $\sim$$3.5\times3.5$ deg$^2$ usually had poor quality. Thus, we used only the central $\sim$$2.5\times2.5$ deg$^2$ regions and applied aggressive sigma clipping to mask bright objects. Note that the diffuse light around the bright objects was not completely masked in practice. Unmasked diffuse light was unlikely to significantly affect the estimation of the global background noise, but we conservatively extended the mask to the neighboring pixels within a radius of 15 pixels to minimize possible contamination from the diffuse light. Figure \ref{fig:fig6} shows histograms of the pixel values of the sky background around NGC 1512. It appeared that the histograms in each band could be described well by a Gaussian distribution once sky values larger than 2$\sigma$ were excluded. 

At the same time, the surface brightness limit can be determined in a variety of ways. For example, it can be measured either directly from 2D images or from an azimuthally averaged radial profile around the target. The latter typically yields a significantly lower (fainter) surface brightness limit than the former. In addition, the surface brightness limit can be estimated from the unbinned original image or from a binned image. In fact, a fainter surface brightness limit can be achieved from a binned image compared with an unbinned one. 

Obtaining azimuthally averaged radial profiles for all galaxies within this data set is beyond the scope of this paper. Thus, we present the surface brightness limit measured from 2D images. However, it would not be practical to measure the limit using pixel-to-pixel variation. Because most of our samples were located within a distance of $\lesssim$20 Mpc, typical LSB objects, such as dwarf satellite galaxies, could be extended to at least $10^{\prime\prime}\times10^{\prime\prime}$ in diameter. This suggested that optimal scaling (binning) was necessary to properly characterize the detection capability of the LSB features. Therefore, referring to the definition described in Appendix A of a study by \cite{2020A&A...644A..42R}, we calculated the 1$\sigma$ surface brightness limit for an angular scale of $10\times10$-arcsec boxes as follows:
\begin{equation}
\mu_{1\sigma; 10^{\prime\prime}\times10^{\prime\prime}} = -2.5\times \mathrm{log}\left(\frac{1\sigma}{\mathrm{pix}\times10^{\prime\prime}}\right)+\mathrm{Zp},
\end{equation}
where pix is the pixel scale of 0.4 arcsec and Zp is the zero point of the data. The zero points of the individual images were estimated using the AAVSO Photometric All-Sky Survey (APASS) DR10 catalog.\footnote{\url{https://www.aavso.org/apass}} As $R$ and $I$ magnitudes are unavailable in the APASS, we computed them using the equations $R=r-0.1837(g-r)-0.0971$ and $I=r-1.2444(r-i)-0.3820$, as provided by Lupton(2005).\footnote{\url{https://www.sdss.org/dr17/algorithms/sdssUBVRITransform/\#Lupton2005}}

As presented in Table \ref{tab:depth}, the calculated surface brightness limit in the broadbands ranged from $\mu_{1\sigma;10^{\prime\prime}\times10^{\prime\prime}}\sim29$ to 31 mag arcsec$^{-2}$ for different filters. Due to the high background level and its Poisson noise, the $I$-band images were the shallowest, as can be identified based on the fact they had the greatest width (shown in Figure \ref{fig:fig6}). Our data set appeared to satisfy the minimum requirements for the surface brightness limit ($\sim$27--28 mag arcsec$^{-2}$) to enable the detection of tidal features and stellar halos \citep{2011ApJ...739...20C,2012arXiv1204.3082B}. In addition, it was deep enough to discover ultra-diffuse galaxies with half-light radii larger than 1.5 kpc and central surface brightnesses fainter than 24 mag arcsec$^{-2}$ \citep{2015ApJ...798L..45V}. 

At the same time, the surface brightness limit would be theoretically correlated with the integration times since background uncertainty can be reduced based on the number of co-added frames. However, practically, the depth may be affected further by a variety of factors, such as the observation conditions, the presence of bright nearby stars, or possibly the strength of diffuse light from Galactic cirrus. 

\begin{figure}[t]
\centering
\includegraphics[width=80mm]{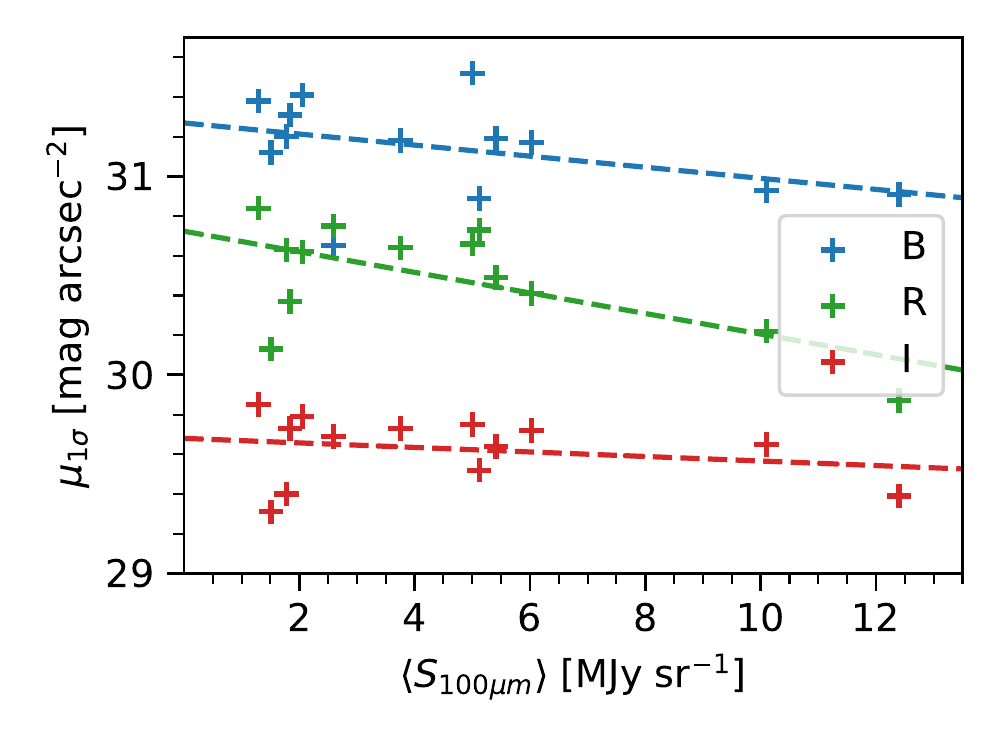}
\caption{Correlation between the average FIR fluxes in the IRAS 100$\mu$m band and the surface brightness limits in the optical bands. Each band is color-coded differently, and the dashed lines represent the results of a linear fit. \label{fig:fig7}}
\end{figure}

Table \ref{tab:depth} provides the galactic coordinates of the sample galaxies and the average FIR fluxes in the sky background within 1 deg$^2$ measured from the IRAS maps. Overall, galaxies with a galactic latitude of $|b| \lesssim 30$ deg were likely to be significantly affected by Galactic cirrus as in the case of $\langle S_{100\mu m}\rangle\gtrsim 2$ MJy sr$^{-2}$. Intriguingly, the surface brightness limit appeared to be slightly correlated with the fluxes of the IRAS 100$\mu$m, especially in the $R$-band (Figure \ref{fig:fig7}). For instance, the correlation coefficient and $p$-value for the $R$-band data were $-0.63$ and 0.02, respectively. This might constitute evidence that Galactic cirrus is bright in the $R$ band and introduces additional uncertainty in the sky estimate. Indeed, this hypothesis is in good agreement with the results of \cite{2020A&A...644A..42R} indicating Galactic cirrus has a bluer $r-i$ color than extragalactic sources for a given $g-r$ color and exhibits a positive correlation between $g-r$ color and IRAS 100$\mu$m flux. 

To calculate the surface brightness limit of the H$\alpha$ emission, additional procedures were required. The H$\alpha$ emission images were created by subtracting the scaled $R$ band images to remove the contribution of the underlying stellar continuum. Field stars were used to determine the relative scale between the H$\alpha$ and $R$ band images. The photometric zero point was also estimated using field stars in the $R$ band image, which may be insensitive to H$\alpha$ emission. Additionally, we recalibrated the measured H$\alpha$ fluxes by accounting for the difference between the H$\alpha$ and $R$-band bandpasses. The details have been presented by \cite{2021ApJ...918...82B}. As a result, the surface brightness limit of H$\alpha$ emission reached $\sim$1--$2\times10^{-18}$ erg s$^{-1}$ cm$^{-2}$ arcsec$^{-2}$. This is slightly deeper than the Local Volume Legacy (LVL) survey data at $\sim$$4\times10^{-18}$ erg s$^{-1}$ cm$^{-2}$ arcsec$^{-2}$ \citep[see][]{2008ApJS..178..247K}. Although our data were not deep enough to detect diffuse H$\alpha$ emission from galaxies \citep[see][]{2016ApJ...817..177L}, they could be useful in analyzing the SF properties of \ion{H}{2} regions, since our survey provides H$\alpha$ images with spatial resolution approximately twice as good as that of the LVL survey.

Although our images have achieved a reasonably satisfactory photometric depth, we do note that they contain significant saturation trails from bright stars (see the figures in the Appendix). These trails can be at least partly mitigated through local interpolation or modeling. However, the degree to which this can be achieved and the level of accuracy necessary depend strongly on the actual science goals. We leave this to the discretion of the user and do not consider it further in this work. 

\subsection{Comparison with other surveys}
It is difficult to quantify the imaging quality of our data set and directly compare it with that from other deep surveys as the definition of the imaging quality strongly depends on the ultimate goals of the surveys. Additionally, the deep surveys have been conducted with a variety of exposure times and telescope sizes, while the methods used to estimate the surface brightness limit also differ. Nevertheless, it is worth comparing the surface brightness limit of our survey with those of other surveys with similar aims to ours. 

\begin{enumerate}[label=(\roman*)]
\item{\cite{2015MNRAS.446..120D} presented deep multi-band images of nearby early-type galaxies obtained with the MegaCam camera at the Canada-France-Hawaii Telescope. The exposure times were typically $\sim$40 min in the $g$, $r$, $i$ bands and twice of that in the $u$ band. They achieved a sky flattening of 0.2\%, yielding a nominal limit of 28.5 mag arcsec$^{-2}$ in the $g$ band as an upper limit.}
\item{\cite{2016ApJ...833..168M} investigated eight nearby Milky Way-analog galaxies as part of the Dragonfly Nearby Galaxies Survey. The total exposure times were typically $\sim$15--20 hr per galaxy. While the 1$\sigma$ surface brightness limits of the co-added $g$ band images were revealed to be 29--30 mag arcsec$^{-2}$ in $60^{\prime\prime}\times60^{\prime\prime}$ boxes, the one-dimensional (1D) surface brightness profile reached 30--32 mag arcsec$^{-2}$ with a 2$\sigma$ depth. After the full 48-lens array was completely built, \cite{2020ApJ...894..119D} achieved 1$\sigma$ depths of 31 mag arcsec$^{-2}$ on arcmin scales with a total exposure time of $\sim$5 hr.}
\item{\cite{2017ApJ...834...16M} conducted a deep imaging survey of galaxies in the Virgo cluster using Case Western Reserve University's 0.6/0.9-m Burrell Schmidt telescope. They obtained a maximum of 100 images each season with 900 sec in the $M$ band for five years and 1,200 sec in the $B$ band for two years. They also obtained a 3$\sigma$ limiting depth of $\mu_B=29.5$ and $\mu_V=28.5$ mag arcsec$^{-2}$ in $60^{\prime\prime}\times60^{\prime\prime}$ boxes.}
\item{The VST Early-type GAlaxy Survey \citep[VEGAS;][]{2015A&A...581A..10C} was designed to obtain deep multi-band photometry in the $g$, $r$, $i$ bands of $\sim$100 nearby galaxies down to $\mu\sim26$--27 mag arcsec$^{-2}$. Further, the surface brightness reached as low as 30 mag arcsec$^{-2}$ using a 1D profile of galaxies calculated from images with a longer total exposure time of 4--5 hr \citep{2020A&A...642A..48I,2021A&A...651A..39R}.}
\item{\cite{2010AJ....140..962M} carried out deep wide-field imaging of eight isolated spiral galaxies in the Local Volume with small (0.1--0.5 m) telescopes. A luminance filter was equipped onto the telescope, and the total exposure times ranged from 5 to 18 hr per galaxy. They obtained a surface brightness limit of $\mu_V\sim28.5$ mag arcsec$^{-2}$ from background rms in boxes, several tens to 100 arcsec per side.}
\item{\cite{2019MNRAS.490.1539R} aimed to recover the halos of $\sim$120 galaxies in the Local Volume using the 0.7-m Jeanne Rich Telescope Centurion 28. With a total exposure time of 1--3 hr, they achieved a surface brightness limit of 28--30 mag arcsec$^{-2}$ from 1D profiles.}
\item{\cite{2021A&A...654A..40T} presented the results of a deep imaging survey exploring the stellar halos and ``missing satellites'' (if any) of nearby galaxies using the $2\times8.4$-m Large Binocular Telescope. The surface brightness limit reached $\mu_V\sim31$ mag arcsec$^{-2}$ (3$\sigma$ in $10^{\prime\prime}\times10^{\prime\prime}$ boxes) with a total exposure time of 2 hr.}
\item{Several surveys mapping the extragalactic sky may be useful in the detection of LSB features. The Hyper Suprime-Cam (HSC) survey \citep{2018PASJ...70S...4A} and Dark Energy Camera Legacy Survey \citep[DECaLS;][]{2019AJ....157..168D} are the most representative and aimed to cover large areas over 1,000 deg$^2$ and 9,000 deg$^2$, respectively. \cite{2021arXiv211103557L} robustly evaluated the depth of these data and suggested that both HSC and DECaLS can achieve a surface brightness limit of $\mu_r\sim28$--29 mag arcsec$^{-2}$ if customized pipelines are applied.}
\item{In addition to measuring the integrated light of galaxies, the direct star count method can be used to study the faint stellar halos of galaxies. \cite{2011ApJS..195...18R} investigated the resolved stellar populations of 14 nearby disk galaxies using the Hubble Space Telescope with a typical exposure time of 700 sec. The median depth was 2.7 mag below the tip of the red giant branch, corresponding to $\mu_V\sim30$ mag arcsec$^{-2}$.}
\end{enumerate}

\section{Scientific Applications} \label{sec:app}
Here, we briefly summarize the potential scientific applications of the KNGS. 

\subsection{Stellar Halos}
Studies on the stellar halos of nearby galaxies have demonstrated that the halo fraction significantly varies from one galaxy to another \citep[e.g.,][]{2016ApJ...833..168M,2019MNRAS.490.1539R}. While stellar halos can be marginally detected with a depth of 28--29 mag arcsec$^{-2}$, 30--31 mag arcsec$^{-2}$ is required for definitive detection \citep[e.g.,][]{2013MNRAS.434.3348C,2016ApJ...833..168M,2021A&A...654A..40T}. Therefore, it may be difficult to explore the photometric properties of stellar halos for all targets in our survey \citep{2018AJ....156..249B}. However, for a part of the sample, one would be able to identify relatively bright substructures, such as stellar streams \citep[e.g.,][]{2010AJ....140..962M,2019ApJ...883L..32V}. This can be used to probe the accretion history of dwarf galaxies and to constrain the gravitational potential of the host galaxy \citep[e.g.,][]{2002MNRAS.332..921I}. 

\subsection{Dwarf Satellite Galaxies}
With the deep images from the KNGS, it is possible to detect satellite galaxies with a central surface brightness of $\mu_R \leq 26$ mag arcsec$^{-2}$ \citep[][and see also \citealt{2022arXiv220300014C}]{2020ApJ...891...18B}. This reveals that the KNGS provides an opportunity to search for new dwarf galaxies in different environments other than the Local Group. \cite{2020ApJ...891...18B} showed that the dwarf galaxies associated with NGC 1291, which resides in a relatively sparse environment, tend to be bluer than those in denser environments. Further, identification of the physical properties of faint companions around XUV-disk galaxies and comparison with non-XUV-disk counterparts would provide useful insights into the physical origins of XUV disks \citep[e.g.,][]{2007ApJS..173..538T}. In combination with knowledge of the distribution of \ion{H}{1} gas, at least for XUV disks in interacting systems, one might be able to detect candidates for tidal dwarf galaxies \citep[TDGs; e.g.,][]{2009MNRAS.400.1749K}. Studying the stellar populations of TDGs with the deep images from the KNGS will help us to explore the nature of TDGs. 

\begin{figure*}[t]
\centering
\includegraphics[width=180mm]{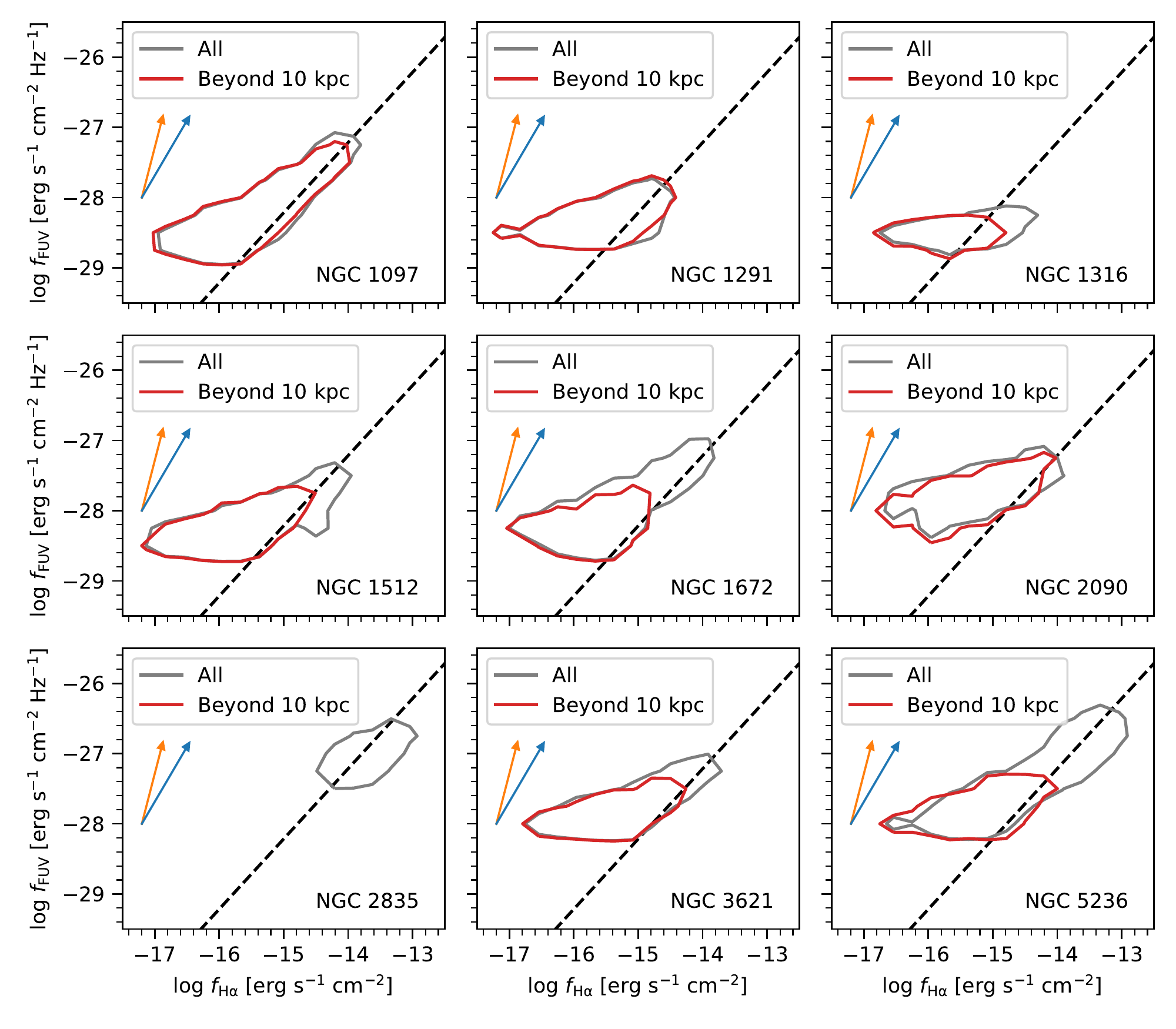}
\caption{Comparison of the FUV flux-density and H$\alpha$ line flux of \ion{H}{2} regions in nine galaxies. All the measurements were corrected for foreground attenuation but not for internal attenuation. The contours contain 90\% of selected data points, and the star-forming regions outside the radial distance of 10 kpc are highlighted with a solid red line. The dashed black line indicates the one-to-one relation between the FUV-inferred SFR and H$\alpha$-inferred SFR. The vectors on each panel show the effect of correction for continuum attenuation of $A_V=1$ with the stellar-to-gas attenuation ratios of 1 (orange) and 0.44 (blue), respectively. \label{fig:fig8}}
\end{figure*}

\subsection{SF Properties in the outer disk}
\cite{2021ApJ...918...82B} demonstrated that broad and narrowband imaging data from the KNGS, along with complementary UV and IR images, are useful in studying the physical properties of star-forming regions on the galaxy outskirts. From an analysis of the spatially resolved spectral energy distribution of NGC 1512 and NGC 2090, \cite{2021ApJ...918...82B} demonstrated that the flux ratio of H$\alpha$ to UV dramatically decreases with galactic radius, indicating that H$\alpha$ flux is deficient in low-mass regions. It appears that the distinctive SF history may be responsible for this trend. Indeed, this result needs to be further explored with other targets from the KNGS. Further, the comparison between XUV and non-XUV-disks in terms of SF among the outskirts will provide valuable information about the physical nature of XUV disks. 

As a preliminary illustration of how the data might be used for a scientific application, we compared the H$\alpha$ and UV fluxes for the nine galaxies that have available far-ultraviolet (FUV) data and detectable H$\alpha$ emission. As described in \cite{2021ApJ...918...82B}, we measured H$\alpha$ and FUV fluxes of star-forming regions in galaxies using an array of circular apertures. Before performing the photometry, we manually masked the galaxy core, foreground stars, and bleeding features. We measured the fluxes with a fixed aperture size of radius 6 arcsec, suitable for covering the physical size of the \ion{H}{2} region at a distance of 10 Mpc.\footnote{Varying aperture sizes for a specific physical scale can be used instead. However, this section aims to determine the FUV and H$\alpha$ flux ratio. In addition, by changing the aperture size, severe aperture correction for distant galaxies may increase uncertainties in the measures.} The primary condition in selecting a star-forming region was whether or not the FUV flux was more than the 3$\sigma$ threshold. Because some artifacts in nearby backgrounds were selected together, we used additional color cuts (e.g., $B-R<1.45$) to effectively discard data that were not considered the galaxy's components. Finally, apertures containing any masked pixels were also excluded. Since investigating the detailed physical properties of each star-forming region is not the purpose of this paper, all the measurements were only corrected for foreground attenuation but not for internal attenuation. 

Figure \ref{fig:fig8} shows a distribution of the FUV flux density and H$\alpha$ flux of the star-forming regions in the nine galaxies. Because we do not yet know the local physical properties of these regions, we focus on the outer regions where dust attenuation is expected to be less significant. Based on the approximate upper limit of the half-light radii of nearby disk galaxies \citep{2011ApJS..197...21H,2019ApJS..244...34G}, we highlight the regions into those that lie exterior to a physical radius of 10 kpc from the full set of regions for the entire galaxy. Most of the galaxies show a similar trend: regions with low flux or low star formation rate (SFR) tend to have deficient H$\alpha$ flux. NGC 2835 is the only exception, because no deep UV data were available and we could not detect any star-forming regions with low flux. These variations of the FUV-to-H$\alpha$ ratio would be difficult to identify using integrated fluxes, highlighting the utility of spatially resolved observations \citep[see also][]{2021ApJ...918...82B}. This suggests that spatially resolved analysis is essential to study the detailed SF properties of galaxies, especially in their outer regions. In future work, we plan to perform spatially resolved SED fitting with available multi-wavelength data to explore the SF properties in XUV- and non-XUV-disk galaxies. 


\begin{acknowledgments}

We are grateful to an anonymous referee for constructive comments and suggestions. This research was supported by the Korea Astronomy and Space Science Institute under the R\&D program(Project No. 2022-1-830-05), supervised by the Ministry of Science and ICT. This research has made use of the KMTNet system operated by the Korea Astronomy and Space Science Institute (KASI), and the data were obtained at CTIO in Chile, SAAO in South Africa, and SSO in Australia. LCH was supported by the National Science Foundation of China (11721303, 11991052, 12011540375) and the China Manned Space Project (CMS-CSST-2021-A04, CMS-CSST-2021-A06). This research was supported by `National Research Council of Science \& Technology (NST)' - `Korea Astronomy and Space Science (KASI)' Postdoctoral Fellowship Program for Young Scientists at KASI in South Korea. This work was supported by the National Research Foundation of Korea (NRF) grant funded by the Korean government (MSIT) (No.2020R1A2C4001753) and under the framework of the international cooperation program managed by the National Research Foundation of Korea (NRF-2020K2A9A2A06026245). Y.K.S. acknowledges support from the National Research Foundation of Korea (NRF) grant funded by the Ministry of Science and ICT (NRF-2019R1C1C1010279). J.H.L. and H.J. were supported by the National Research Foundation of Korea(NRF) grant funded by the Korea government(MSIT) (No. 2022R1A2C1004025).

\end{acknowledgments}

%

\vspace{5mm}
\facilities{KMTNet, AAVSO, IRSA, NED}


\software{Astropy \citep{2013A&A...558A..33A,2018AJ....156..123A}, Scipy \citep{2020SciPy-NMeth}, Source Extractor \citep{1996A&AS..117..393B}, Photutils \citep{2021zndo...4624996B}, SCAMP \citep{2006ASPC..351..112B}, SWarp \citep{2002ASPC..281..228B}, Matplotlib \citep{Hunter:2007}}



\bibliography{ms}{}
\bibliographystyle{aasjournal}

\appendix

\section{Co-added images of the samples with a full FoV}
Section \ref{sec:iq} provides the surface brightness limits calculated from the background noise of individual images for a quantitative assessment of the imaging quality of the data set. At the same time, we need to be careful about accepting the limits at face value. Our images contain a variety of artifacts, such as saturation trails, crosstalks, and stray light from stars. In particular, the last two elements may be mistakenly considered as LSB features or overlapped with actual LSB features. This means that the completeness and reliability of the LSB detection may vary for each image regardless of the surface brightness limit. Therefore, a subsequent visual assessment is necessary to understand the imaging quality of the data. For this purpose, we also present final mosaic images of all the samples. We also provide H$\alpha$ emission maps overlaid on the $R$-band images to highlight the spatial distribution of star-forming regions. The nuclear regions in some galaxies might be affected by imperfect continuum subtraction. 

\begin{figure*}[p]
\centering
\includegraphics[width=130mm]{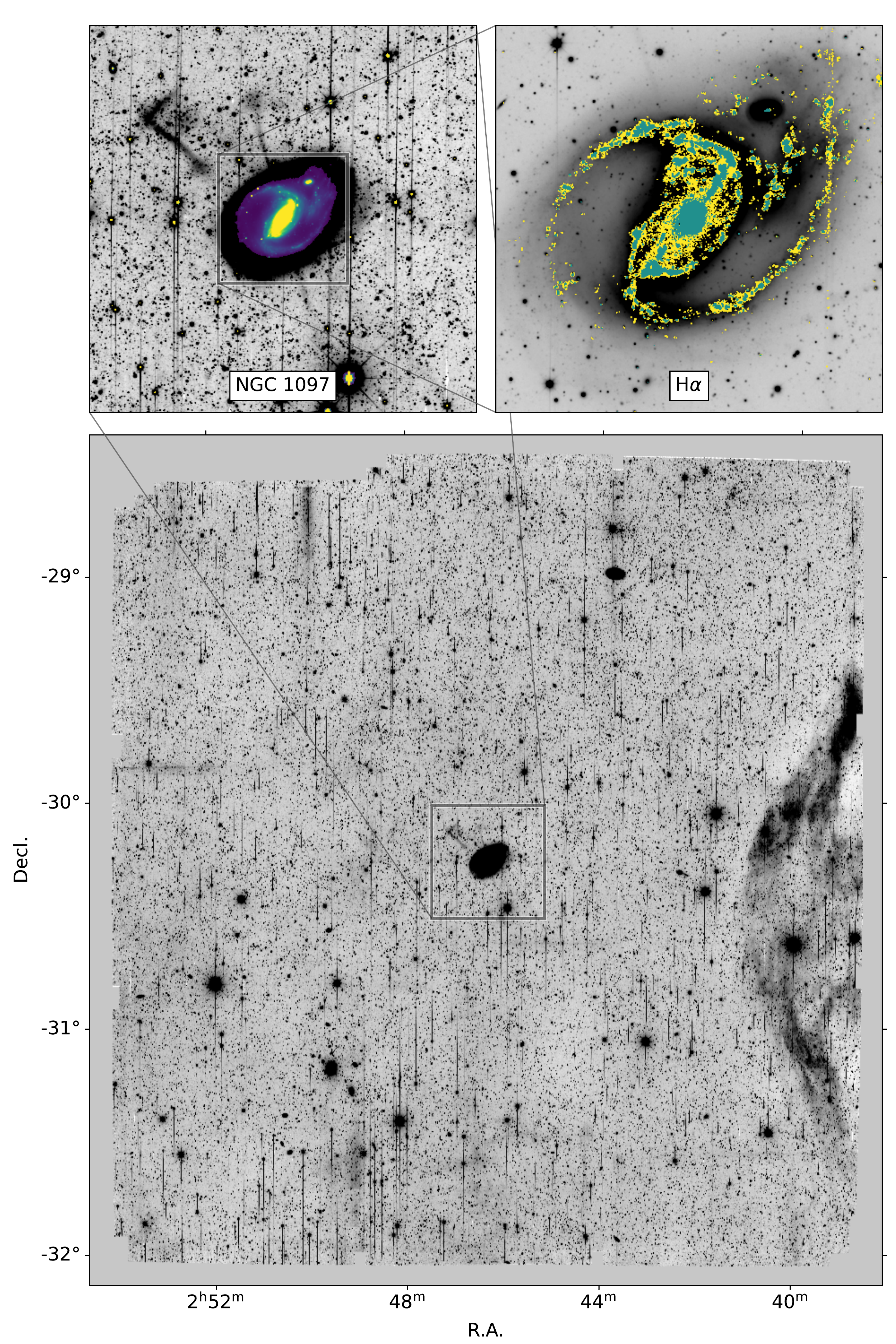}
\caption{Final mosaic image of NGC 1097 in the $R$ band. The upper panels show zoomed-in images with a FoV of $30^\prime\times30^\prime$ (left) and $10^\prime\times10^\prime$ (right), respectively. The color-coded regions in the upper-left panel represent the structures with a surface brightness of $\mu_R\lesssim25$ mag arcsec$^{-2}$, while the ones in the upper-right panel depict regions with H$\alpha$ flux levels 1$\sigma$ (yellow) and 10$\sigma$ (green) above the background. \label{fig:figA-1}}
\end{figure*}

\begin{figure*}[p]
\centering
\includegraphics[width=130mm]{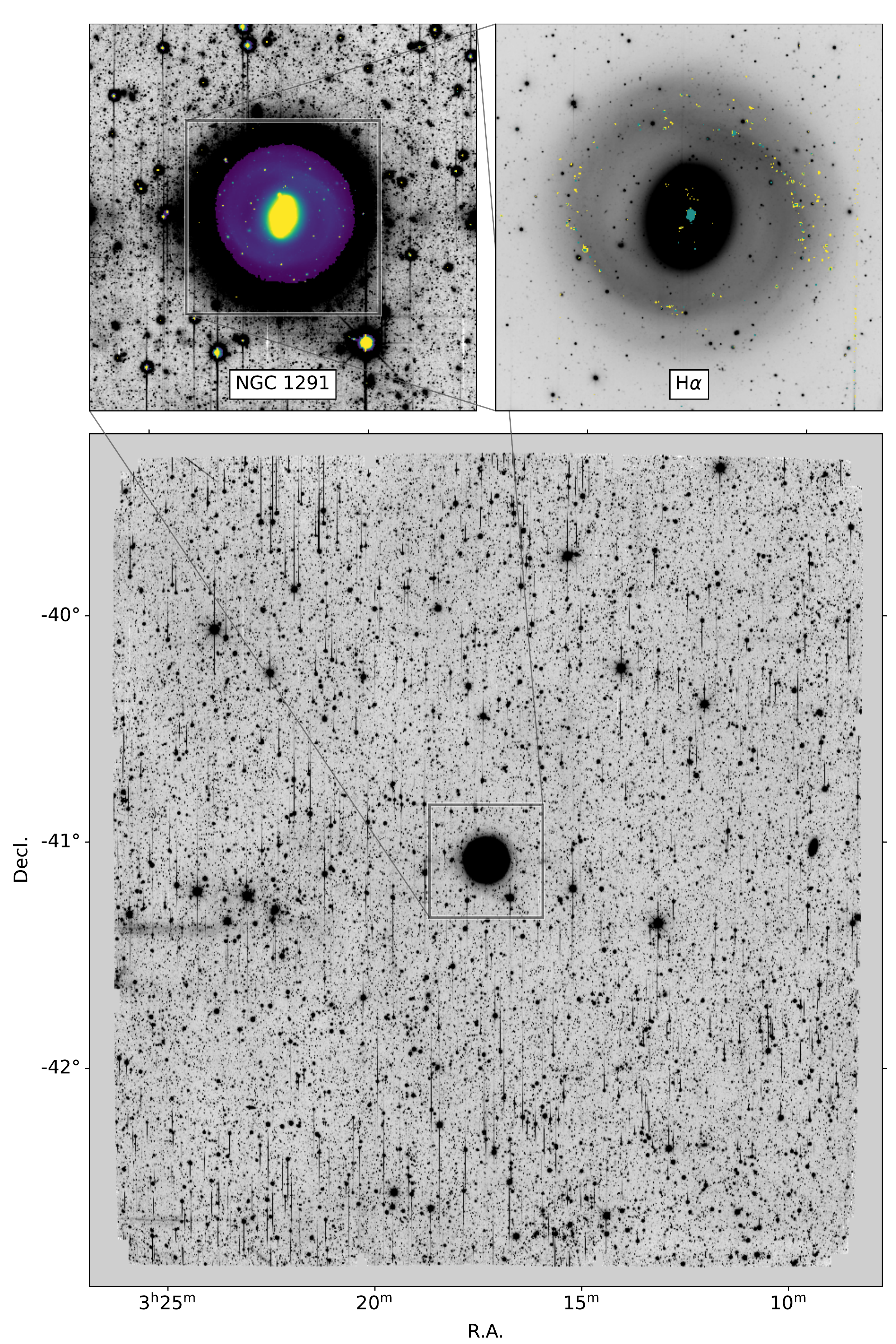}
\caption{Final mosaic image of NGC 1291 in the $R$ band. The other descriptions are the same as in Figure \ref{fig:figA-1}, except that the FoV of the upper-right panel is $15^\prime\times15^\prime$.} \label{fig:figA-2}
\end{figure*}

\begin{figure*}[p]
\centering
\includegraphics[width=130mm]{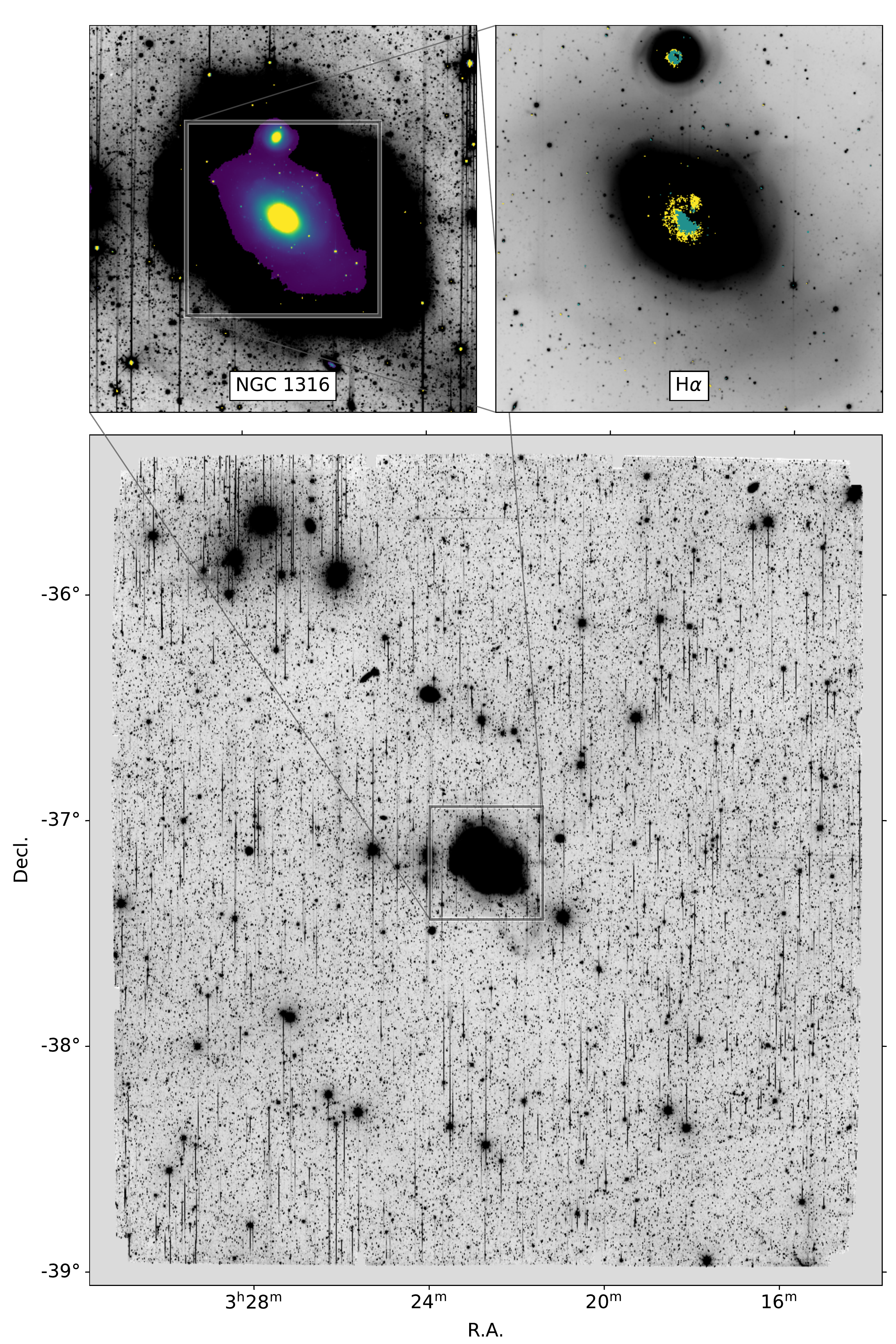}
\caption{Final mosaic image of NGC 1316 in the $R$ band. The other descriptions are the same as in Figure \ref{fig:figA-1}, except that the FoV of the upper right panel is $15^\prime\times15^\prime$.} \label{fig:figA-3}
\end{figure*}

\begin{figure*}[p]
\centering
\includegraphics[width=130mm]{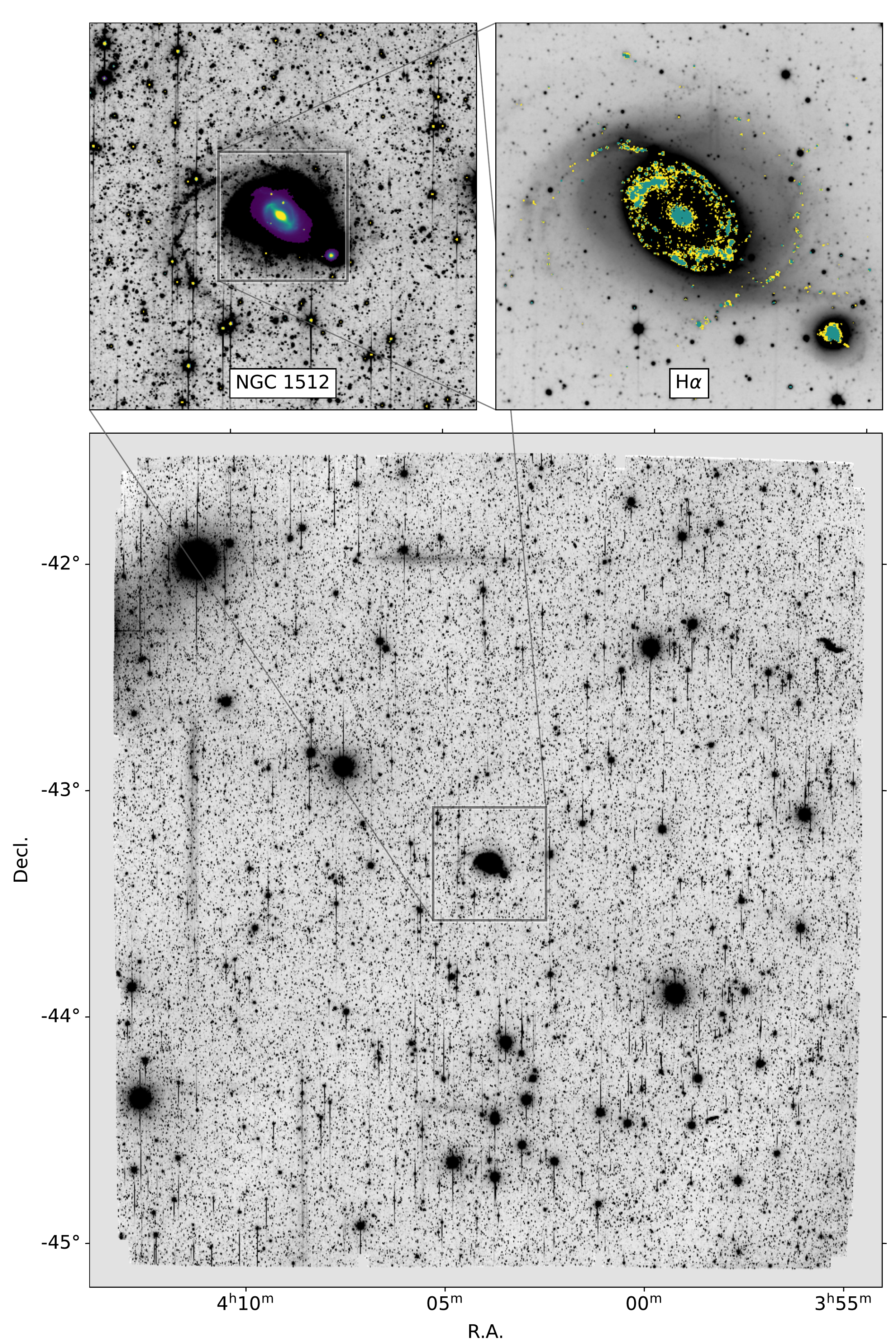}
\caption{Final mosaic image of NGC 1512 in the $R$ band. The other descriptions are the same as in Figure \ref{fig:figA-1}.} \label{fig:figA-4}
\end{figure*}

\begin{figure*}[p]
\centering
\includegraphics[width=130mm]{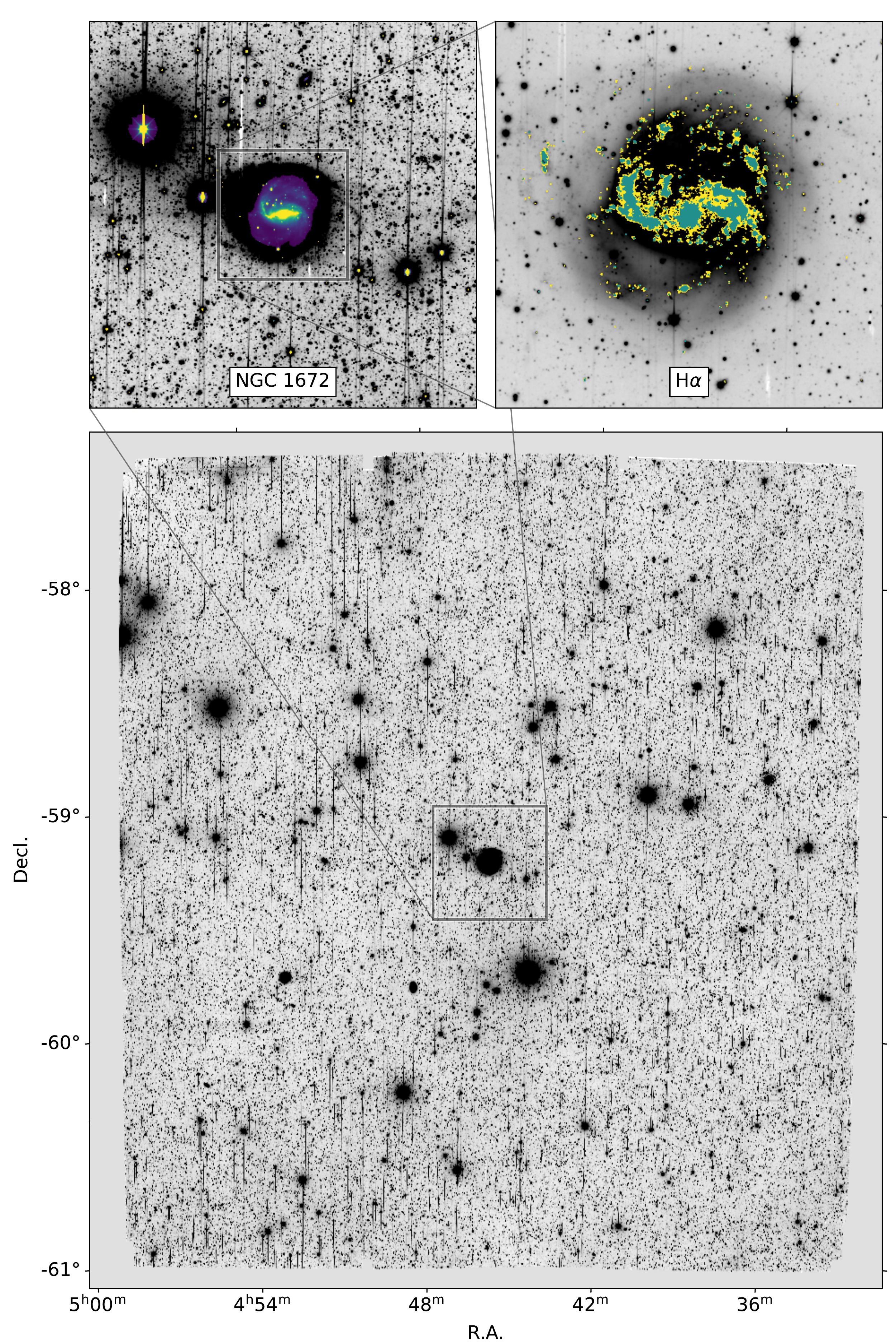}
\caption{Final mosaic image of NGC 1672 in the $R$ band. The other descriptions are the same as in Figure \ref{fig:figA-1}.} \label{fig:figA-5}
\end{figure*}

\begin{figure*}[p]
\centering
\includegraphics[width=130mm]{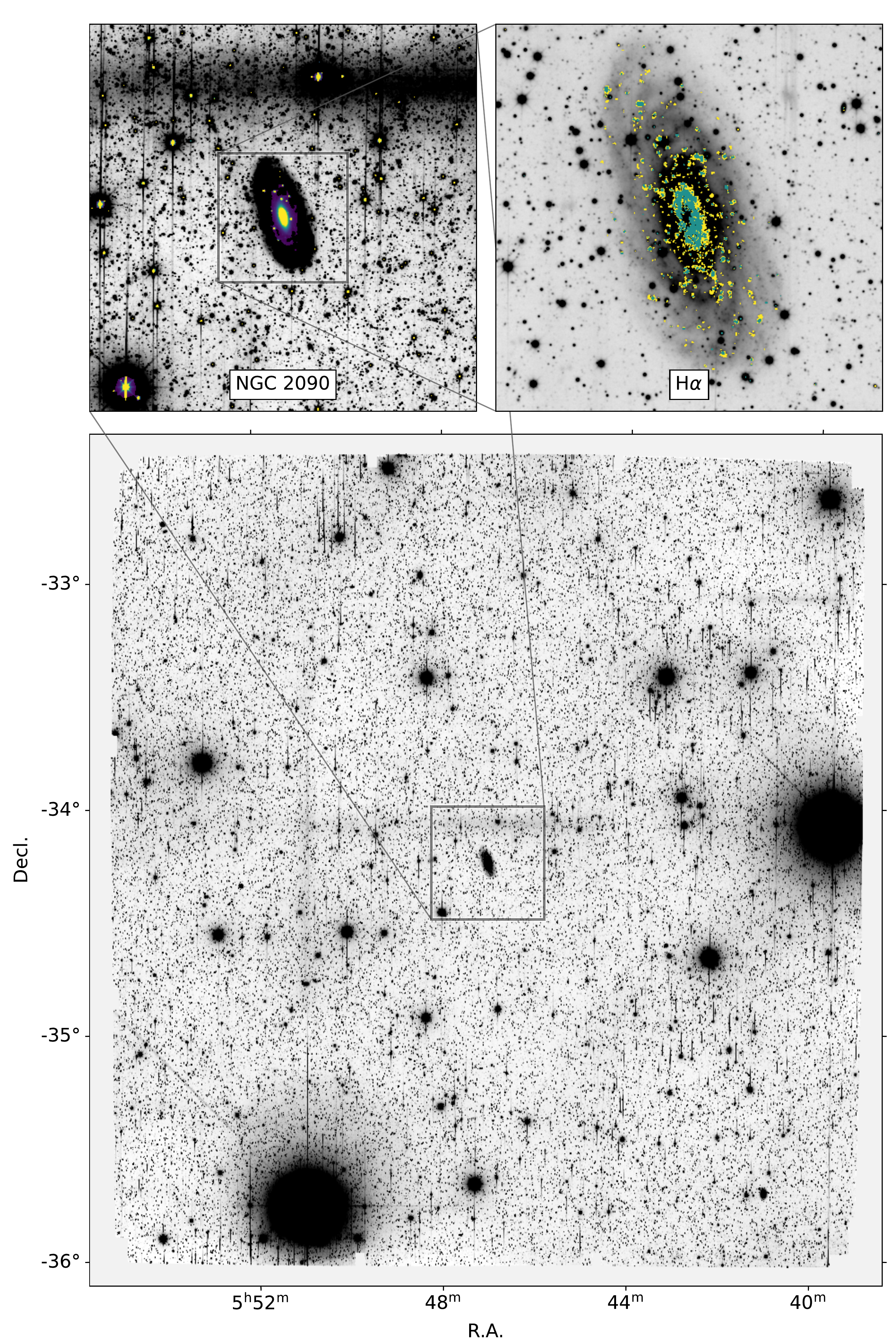}
\caption{Final mosaic image of NGC 2090 in the $R$ band. The other descriptions are the same as in Figure \ref{fig:figA-1}.} \label{fig:figA-6}
\end{figure*}

\begin{figure*}[p]
\centering
\includegraphics[width=130mm]{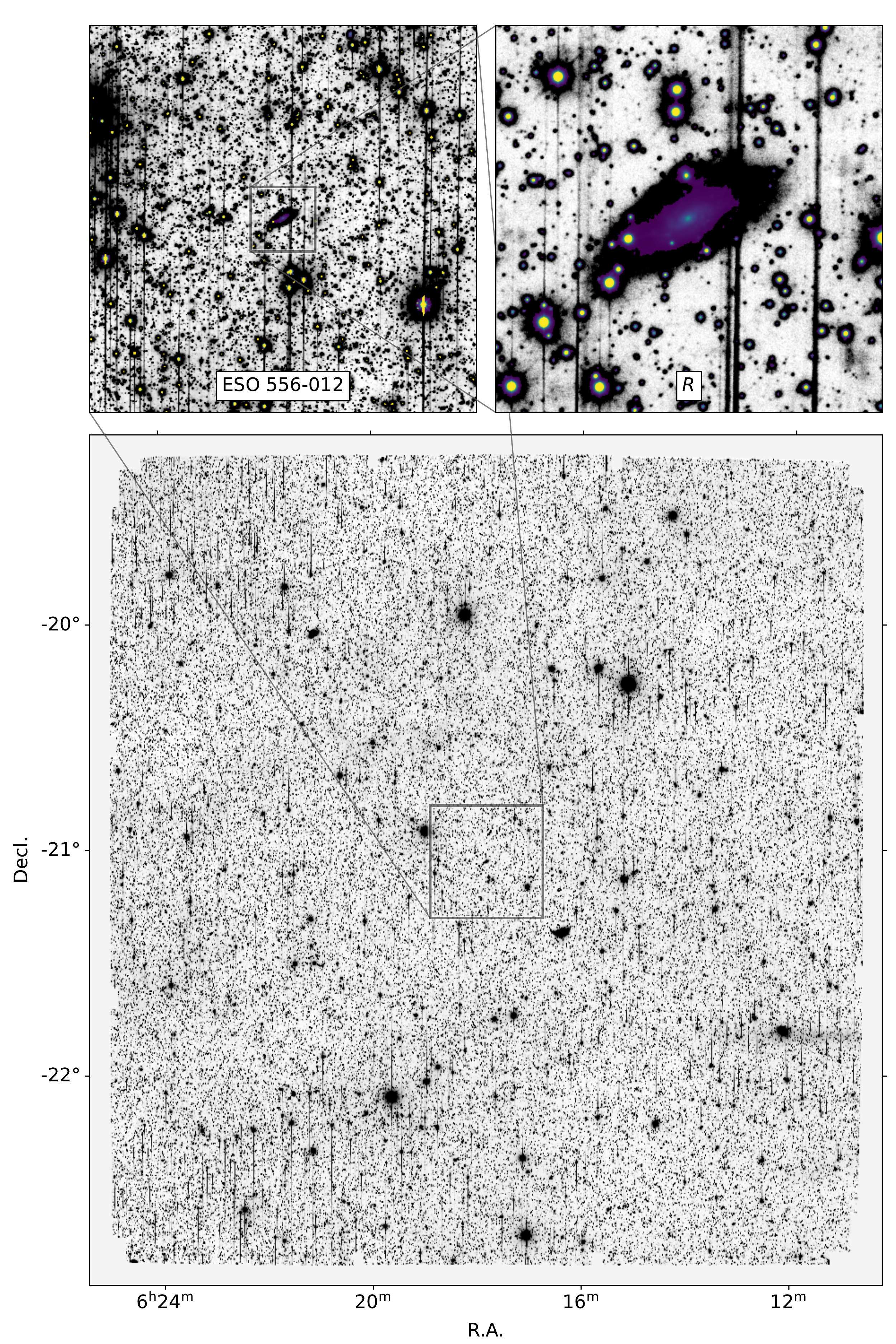}
\caption{Final mosaic image of ESO 556-012 in the $R$ band. The other descriptions are the same as in Figure \ref{fig:figA-1}, except that the upper-right panel shows a zoomed-in $R$-band image with a FoV of $5^\prime\times5^\prime$.} \label{fig:figA-7}
\end{figure*}

\begin{figure*}[p]
\centering
\includegraphics[width=130mm]{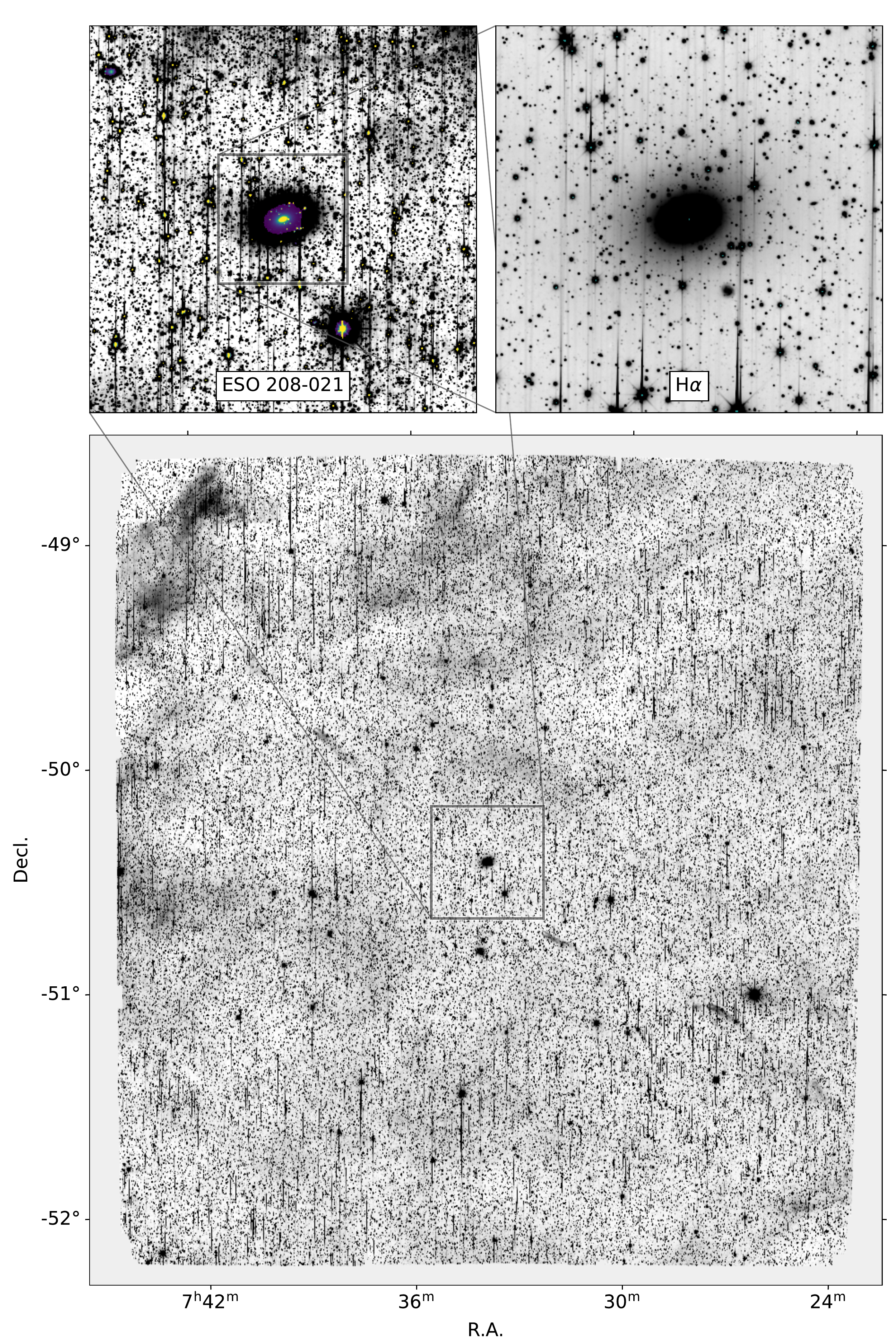}
\caption{Final mosaic image of ESO 208-021 in the $R$ band. The other descriptions are the same as in Figure \ref{fig:figA-1}.} \label{fig:figA-8}
\end{figure*}

\begin{figure*}[p]
\centering
\includegraphics[width=130mm]{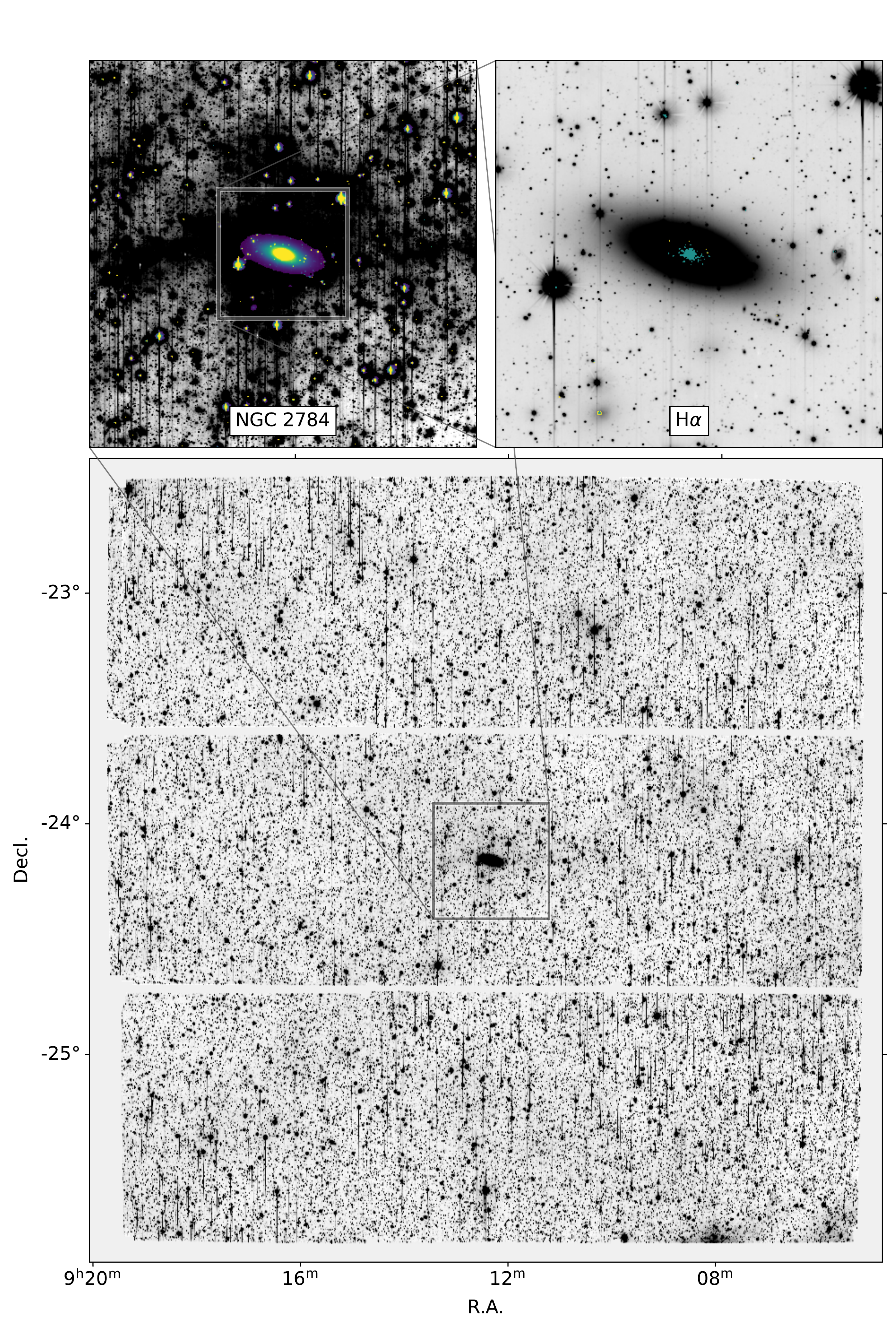}
\caption{Final mosaic image of NGC 2784 in the $R$ band. The other descriptions are the same as in Figure \ref{fig:figA-1}.} \label{fig:figA-9}
\end{figure*}

\begin{figure*}[p]
\centering
\includegraphics[width=130mm]{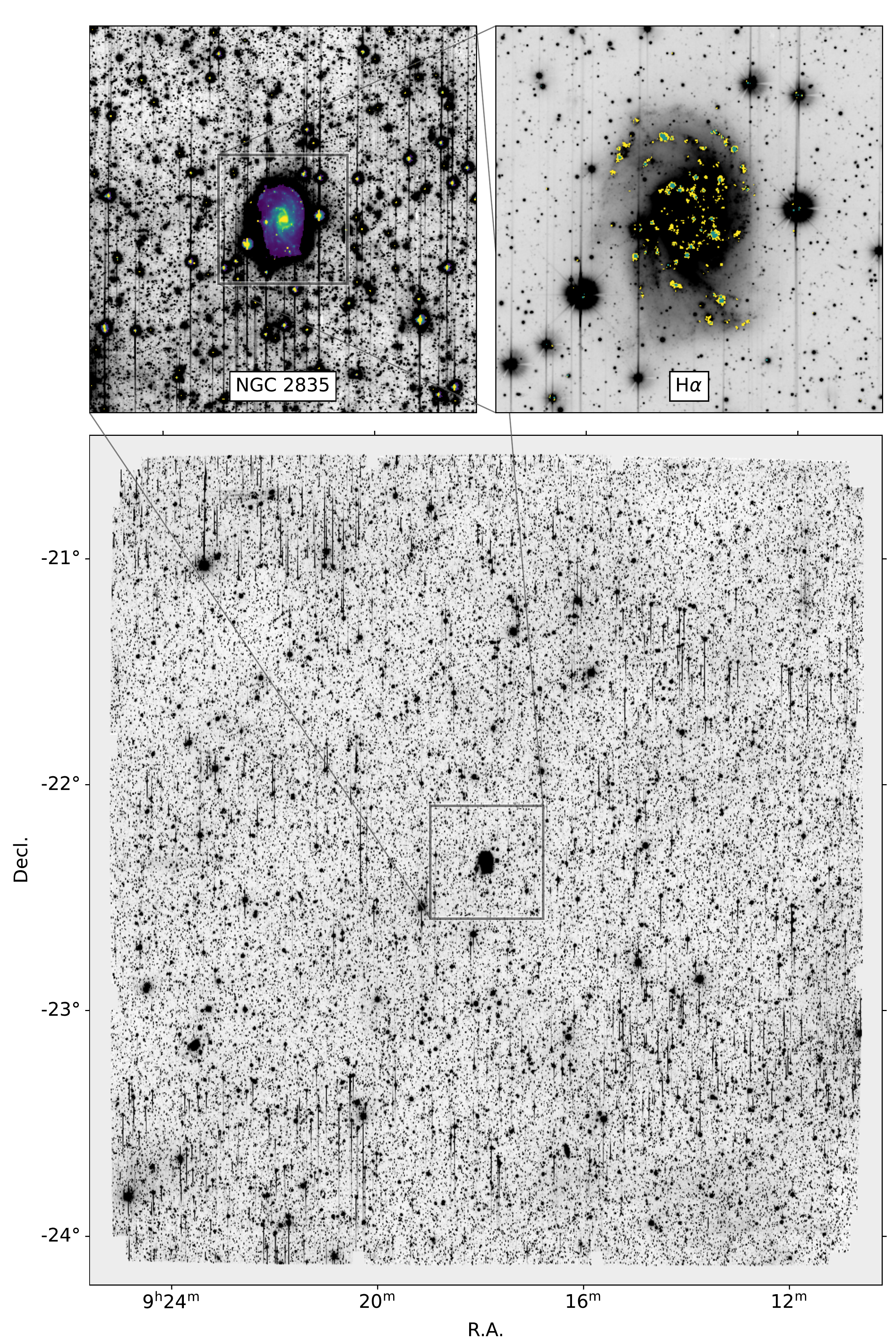}
\caption{Final mosaic image of NGC 2835 in the $R$ band. The other descriptions are the same as in Figure \ref{fig:figA-1}.} \label{fig:figA-10}
\end{figure*}

\begin{figure*}[p]
\centering
\includegraphics[width=130mm]{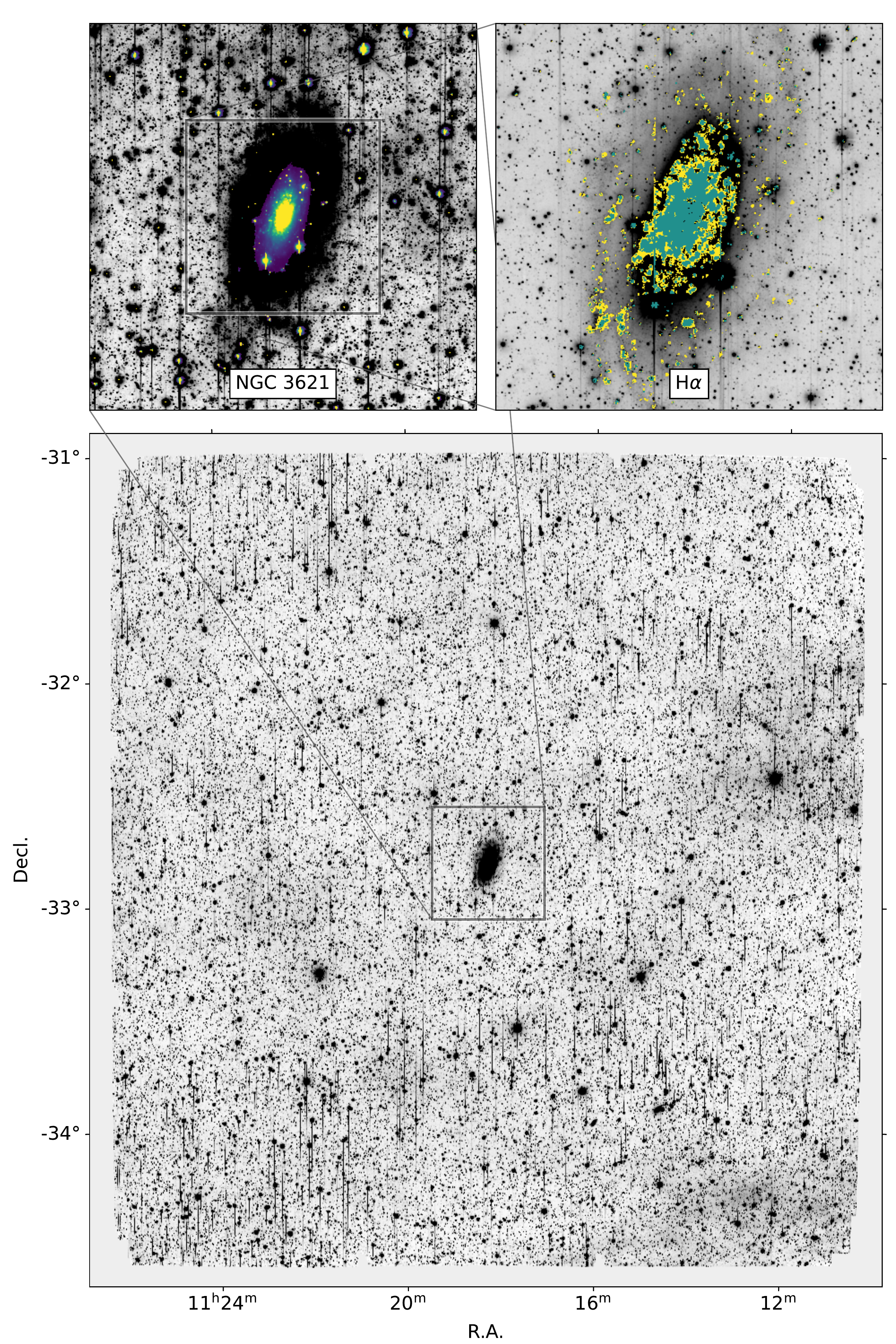}
\caption{Final mosaic image of NGC 3621 in the $R$ band. The other descriptions are the same as in Figure \ref{fig:figA-1}, except that the FoV of the upper right panel is $15^\prime\times15^\prime$.} \label{fig:figA-11}
\end{figure*}

\begin{figure*}[p]
\centering
\includegraphics[width=130mm]{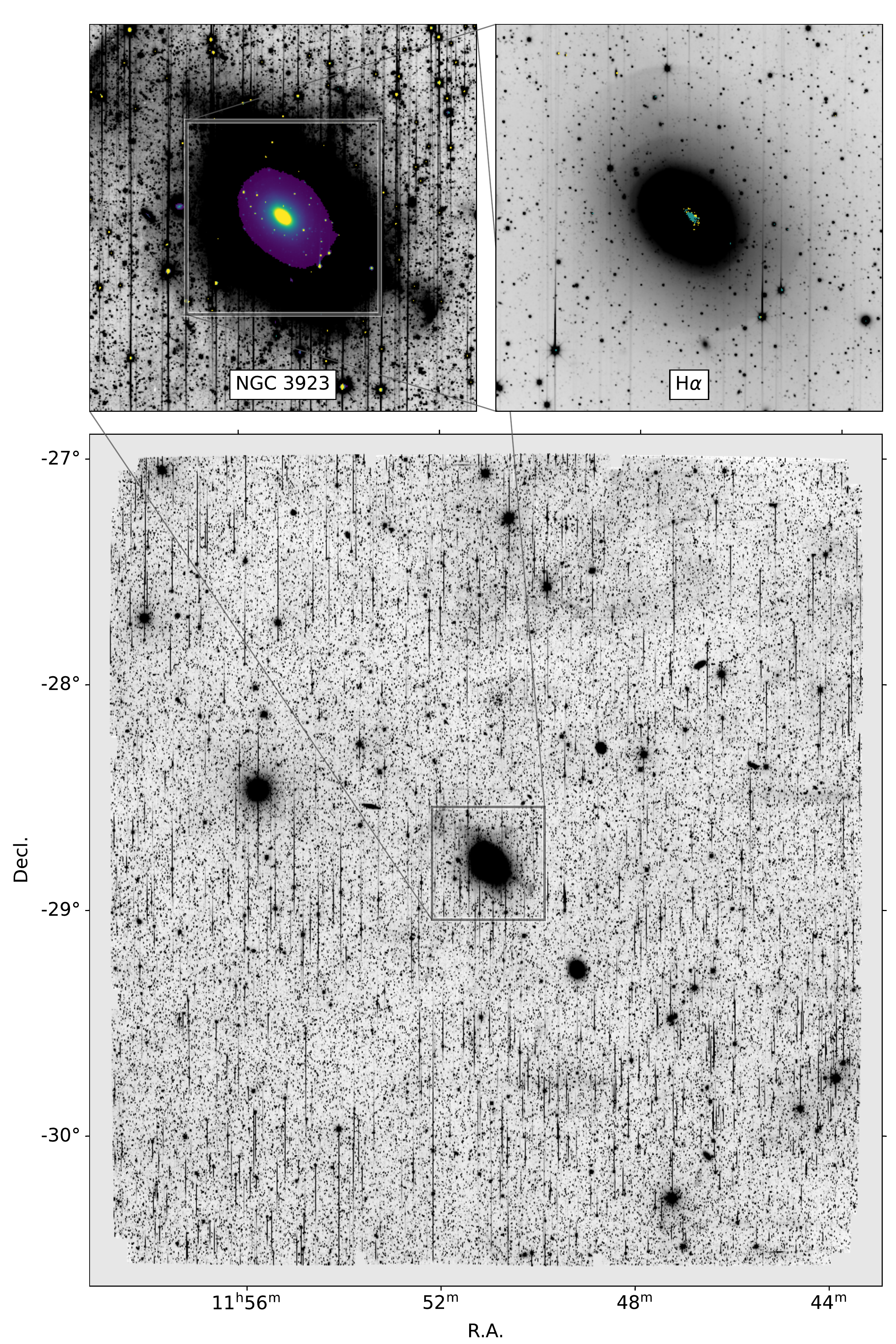}
\caption{Final mosaic image of NGC 3923 in the $R$ band. The other descriptions are the same as in Figure \ref{fig:figA-1}, except that the FoV of the upper right panel is $15^\prime\times15^\prime$.} \label{fig:figA-12}
\end{figure*}

\begin{figure*}[p]
\centering
\includegraphics[width=130mm]{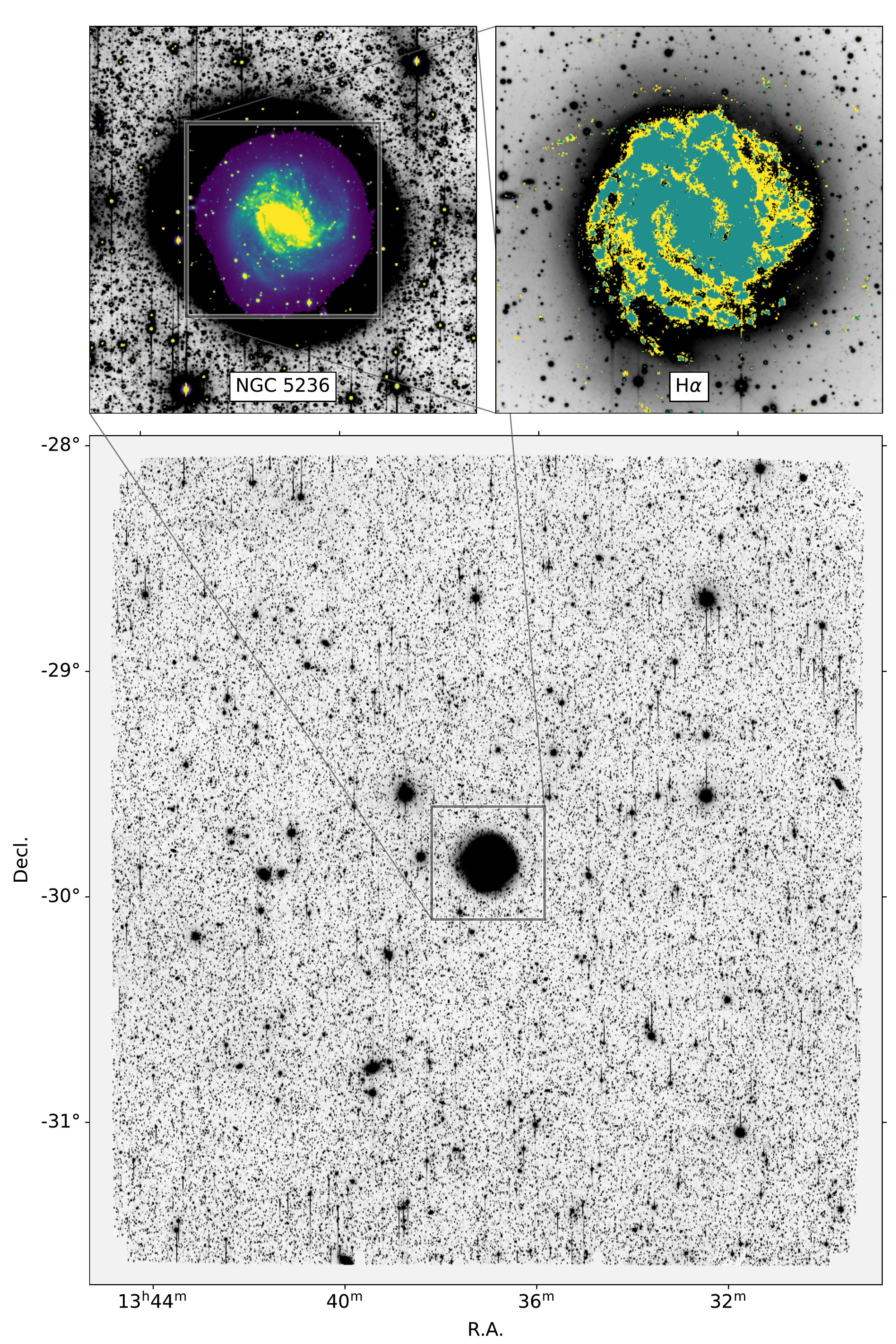}
\caption{Final mosaic image of NGC 5236 in the $R$ band. The other descriptions are the same as in Figure \ref{fig:figA-1}, except that the FoV of the upper right panel is $15^\prime\times15^\prime$.} \label{fig:figA-13}
\end{figure*}





\end{document}